\documentclass[twocolumn,amsmath,amssymb,pra,superscriptaddress]{revtex4}

\pdfoutput=1

\usepackage{graphicx}
\usepackage{dcolumn}
\usepackage{bm}
\usepackage{braket}
\usepackage{import}
\usepackage{comment}
\usepackage{framed}
\usepackage{tikz}
\usepackage{calrsfs}
\usepackage{dsfont}
\usepackage{adjustbox}
\usepackage{graphicx}
\usepackage[colorlinks=true,citecolor=blue,urlcolor=purple]{hyperref}

\newcommand{\moy}[1]{\langle #1 \rangle}
\newcommand{\pmat}[4]{\begin{pmatrix} #1 & #2 \\ #3 & #4\end{pmatrix}}
\DeclareMathOperator{\Tr}{Tr}

\begin{document}

\title{Implementable Hybrid Entanglement Witness}

\author{Ga\"el Mass\'e}
\email{gael.masse@univ-paris-diderot.fr}
\affiliation{Universit\'e de Paris, Laboratoire Mat\'eriaux et Ph\'enom\`enes Quantiques, CNRS, F-75013, Paris, France}
\author{Thomas Coudreau}
\affiliation{Universit\'e de Paris, Laboratoire Mat\'eriaux et Ph\'enom\`enes Quantiques, CNRS, F-75013, Paris, France}
\author{Arne Keller}
\affiliation{Universit\'e Paris-Saclay, Laboratoire Mat\'eriaux et Ph\'enom\`enes Quantiques, CNRS, F-75013, Paris, France}
\author{Perola Milman}
\affiliation{Universit\'e de Paris, Laboratoire Mat\'eriaux et Ph\'enom\`enes Quantiques, CNRS, F-75013, Paris, France}

\date{\today}

\begin{abstract}
  Hybrid encoding of quantum information is a promising approach towards the realisation of optical quantum protocols. It combines advantages of continuous variables encoding, such as high efficiencies, with those of discrete variables, such as high fidelities. In particular, entangled hybrid states were shown to be a valuable ressource for quantum information protocols. In this work, we present a hybrid entanglement witness that can be implemented on currently available experiments and is robust to noise currently observed in quantum optical set-ups. The proposed witness is based on measurements of genuinely hybrid observables. The noise model we consider is general. It is formally characterised with Kraus operators since the considered hybrid system can be expressed in a finite dimension basis. A practical advantage of the witness is that it can be tested by measuring just a few experimentally available  observables. 
\end{abstract}

\maketitle

\section{Introduction}
Many different platforms are envisaged to process quantum information, corresponding to different ways of encoding qubits. All these implementations fall into
two main categories: discrete variables (DV), based on observables with discrete spectra and continuous variables (CV), based on ones with continuous spectra.
Both regimes present specific advantages and drawbacks: while DV show high fidelities, their efficiencies are in general low and the contrary 
applies for CV implementations \cite{van2011optical, qi_nonlinear_2016}. Hybridization between DV and CV states can take advantage of both encodings to implement certain quantum protocols \cite{takeda2019toward}. An example is near deterministic teleportation with high 
fidelities \cite{takeda_deterministic_2013, lee_near-deterministic_2013, lie2019limitations}, steering \cite{PhysRevLett.121.170403}, Bell protocols \cite{brask2012bell, quintino2012maximal, kwon2013violation,T_ppel_2015} and hybrid quantum repeaters \cite{brask2010hybrid, PhysRevA.99.032349Repeaters}. Quantum information processing using this technique is currently being developed both theoretically 
\cite{lim_loss-resilient_2016, andersen_hybrid_2015, kwon_generation_2015} and 
experimentally \cite{engineeringopticalhybrid, PhysRevLett.121.170403, guccione2020, sychev2018, gouzien2020hybrid}. 

Entanglement lies at the heart of quantum physics and is a key resource for 
quantum information and computation \cite{RevModPhys.91.025001, 
horodecki2009quantum}. Its detection is thus of crucial importance 
and has been studied extensively, notably with so-called entanglement 
witnesses (EW) \cite{horodecki2009quantum}. The fact that there exist EW for every entangled state \cite{horodecki_separability_1996} has raised their importance on a
theoretical point of view even further \cite{chruscinski_entanglement_2014}, and links between entanglement witness and other 
important features of quantum physics such as Bell inequalities have been assessed
\cite{hyllus_relations_2005}. Whenever one is interested in hybrid ressource, the issue of entanglement appears naturally, since we deal with a bipartite quantum system. As a consequence, the complementarity principle will involve producing entangled states. For this reason, entanglement detection is a foundational issue in hybrid encoding.

Entanglement witnesses (EW) have been studied extensively for discrete
\cite{guhne_entanglement_2009} and continuous \cite{sperling_necessary_2009} systems. Nevertheless, EW involving measurements of observables with a continuous spectrum seem
harder to establish \cite{ qi_nonlinear_2016}. This is particularly true if the states considered are non-gaussian, which is precisely the 
case of all hybrid states \cite{kreis_classifying_2012}. The complete knowledge of the system's density matrix is a sufficient condition to compute EWs, ~\cite{peres_separability_1996,simon_peres-horodecki_2000,arkhipov2018negativity,hou_constructing_2010,
guo_sufficient_2011}, but it is not necessary. Besides, this is not a practical solution since it requires time demanding quantum tomography techniques.

One natural way for obtaining entanglement witness in CV systems is to use 
inseparability criteria based 
on matrices of moments \cite{miranowicz_inseparability_2009, gittsovich_non-classicality_2015}, an 
approach subsumed in Ref.~\cite{PhysRevLett.95.230502} and
applied in~Refs.~\cite{PhysRevLett.84.2726, PhysRevLett.84.2722, 
  PhysRevLett.88.120401, 
PhysRevA.67.052104, PhysRevLett.96.050503}, which can be generalised to hybrid system \cite{van2011optical}. Another approach was given in 
\cite{arkhipov2018negativity} where it was shown that the negativity 
volume of the generalised Wigner function can be used to detect entanglement for hybrid states. These approaches are however too sensitive to noise or too costly in terms of measurements with regard to our goals.

In this work we introduce an implementable entanglement witness 
on a given quantum optics setup where
hybrid entangled states are currently produced experimentally \cite{van2011optical, morin_remote_2014}.
Our approach is inspired by the well known entanglement witness \cite{chruscinski_entanglement_2014}
\begin{equation}
  W = \lambda\mathds{1} - \ket{\psi}\bra{\psi}
\end{equation}
where $\lambda \in\mathbb{R}$ is optimised such that 
$\Tr[W\sigma] >0$ for any separable state $\sigma$ and
$\Tr[W\rho]<0$, for the largest possible set of entangled states including $\rho=\ket{\psi}\bra{\psi}$. We then adapt
$W$ so that it is robust to noise using a realistic noise model, and require the measurement of only a few observables. We choose to stick to a specific experimental setup to produce a concrete and experimentally realistic example of an efficient hybrid entanglement detection. However, the construction of the witness enables its adaptation to other experimental platforms using different encodings, as for instance in \cite{gouzien2020hybrid}, as we will show.

After introducing the set-up, we analyse the evolution of the hybrid entangled state of interest under a general noise 
model (II) and show that we can define a suitable witness using only measurable genuinely hybrid observables (III). We then discuss the efficiency of the introduced witness (IV) before concluding (V). 

\section{\label{stateanalysis} The setup }
We start by introducing the family of entangled states we aim to characterise.
\subsection{The target states}
We consider the experimental quantum optics set-up described in details in
Ref.~\cite{morin_remote_2014}. It is designed to produce, in the ideal scenario, the
following pure state of the electromagnetic field: 
\begin{equation} 
  \ket{\psi} = \frac{\ket{0} \ket{C^{-}(\alpha)} + \ket{1} \ket{C^{+}(\alpha) 
  }}{\sqrt{2}},
  \label{HybridEntangledState}
\end{equation}
where
\begin{equation}
  \ket{C^{\pm}(\alpha)} = \frac{\ket{\alpha} \pm \ket{- 
  \alpha}}{N^{\pm}(\alpha)}
  \label{eq:catStateDef}
\end{equation}
are the so-called symmetric and antisymmetric ``Schr\"odinger cat''-like 
states, with $\ket{\alpha}$ being a coherent state of 
amplitude $\alpha$, and $N^{\pm}(\alpha) = 2(1\pm \Re[e^{-2\alpha}])$, so that 
$\braket{\psi|\psi}=1$.
Its specific advantage with respect to the hybrid state $\frac{\ket{0} 
\ket{\alpha}+ \ket{1}\ket{- \alpha}}{\sqrt{2}}$, which was considered 
in~\cite{van2011optical,jeong_generation_2014}, is that in Eq.~\eqref{eq:catStateDef} the two considered continuous variables states are 
orthogonal to each other for all values of $\alpha$. From now on, $\alpha$ will be taken real, without loss of generality. As for the discrete part of $\ket{\psi}$, we consider, as in 
Ref.~\cite{morin_remote_2014}, that $\ket{0}$ is the vacuum and $\ket{1}$ is 
the Fock state with one photon. However, the derivation of the witness that we present here can be adapted to other 
discrete encodings such as orthogonal polarization states of the photon \cite{kwon_generation_2015, FangPola}.

In experiments, the produced state is noisy and should be described by a density matrix 
$\rho_{\text{noise}}$ instead of $\ket{\psi}$. A correct modelisation of $\rho_{\text{noise}}$ depends
crucially on the type of encoding as well as on the specificities of the considered experimental setup. In the present context, we
consider photon losses in both discrete and continuous channel as being the main source of noise. 
Such losses can be modelled by the action of a beam-splitter (BS) ~\cite{Leonhardt1993} which
entangles an ideal incoming state $\ket{\psi}\bra{\psi}$ with an 
ancillary fluctuating quantum field. We note $\rho_{da}$ and $\rho_{ca}$ the ancillary fields, respectively on the discrete channel and the continuous channel, and these beam-splitters are refered to as TBS, for theoretical beam-splitters, in the scheme we propose in Fig.~\ref{schemeofsetupvacuum}. After recombination on the beam-splitter, two outputs are produced corresponding to the transmitted part of the beam-splitter and to the reflected one. We trace out the reflected one which corresponds to the losses, and obtain the mixed state $\rho_{\text{noise}} = \Tr_{\text{r}_{\text{c}},\text{r}_{\text{d}}}[\ket{\psi_{\text{noise}}}\bra{\psi_{\text{noise}}}]$ with 
\begin{align}
  &\ket{\psi_{\text{noise}}} \nonumber \\
  &= \frac{\left( \ket{\sqrt{1\text{-}\eta}\alpha}_{\text{t}_{\text{c}}} \ket{\sqrt{\eta}\alpha}_{\text{t}_{\text{c}}} - \ket{\text{-}
  \sqrt{1\text{-}\eta}\alpha}_{\text{t}_{\text{c}}} \ket{\text{-}\sqrt{\eta}\alpha}_{\text{t}_{\text{c}}}  \right) \ket{0}_{\text{t}_{\text{d}}} \ket{0}_{\text{r}_{\text{d}}}}{\sqrt{2}N^{-}(\alpha)} \nonumber
  \\
  &+ \frac{ \left( 
      \ket{\sqrt{1\text{-}\eta}\alpha}_{\text{t}_{\text{c}}} \ket{\sqrt{\eta}\alpha}_{\text{t}_{\text{c}}}\footnotesize{+}\ket{\text{-} 
  \sqrt{1\text{-}\eta}\alpha}_{\text{t}_{\text{c}}} \ket{\text{-}\sqrt{\eta}\alpha}_{\text{t}_{\text{c}}}  \right) \sqrt{1\text{-}\eta_d}\ket{1}_{\text{t}_{\text{d}}} \ket{0}_{\text{r}_{\text{d}}} }{\sqrt{2}N^{+}(\alpha)}  \nonumber  \\
  &+ \frac{ \left( 
      \ket{\sqrt{1\text{-}\eta}\alpha}_{\text{t}_{\text{c}}} \ket{\sqrt{\eta}\alpha}_{\text{t}_{\text{c}}}+\ket{\text{-}
  \sqrt{1\text{-}\eta}\alpha}_{\text{t}_{\text{c}}} \ket{\text{-}\sqrt{\eta}\alpha}_{\text{t}_{\text{c}}}  \right)  \sqrt{\eta_d}\ket{0}_{\text{t}_{\text{d}}} \ket{1}_{\text{r}_{\text{d}}} }{\sqrt{2}N^{+}(\alpha)} 
\end{align}

where $\Tr_{\text{r}_{\text{c}},\text{r}_{\text{d}}}$ 
denotes the partial trace over the reflected modes, respectively in the continuous and discrete channels, $\text{t}_{\text{c}}$ and $ \text{t}_\text{d}$ are the transmitted modes respectively in the continuous and discrete channels and
$\eta^2, \eta_d^2$ are the reflexivity of the theoretical beam-splitters, respectively for the continuous channel and for the discete channel. Therefore, $\eta$ and $\eta_d \in [0,1]$ 
characterise the noise in both channels $\eta_{(d)} = 0$ being 
the ideal case and $\eta_{(d)} =1$ the completely noisy channel.

\begin{figure}
  \begin{tikzpicture}[thick, scale=0.36]
    \draw(10,1)--(10,2) ; 
    \draw(0,2)--(12,2) ; 
    \draw(9,1)--(11,3) ; 
    \draw(7,3)--(5,1) ; 
    \draw(6,2)--(6,8) ; 
    \draw(6,5)--(7,5) ; 
    \draw(0,8)--(12,8) ; 
    \draw(5,9)--(7,7) ; 
    \draw(9,9)--(11,7) ; 
    \draw(10,9)--(10,8) ; 
    \draw (10,1) node[below]{$\rho_{ca}$} ;
    \draw (0,2) node[below]{Photon Pair} ;
    \draw (12,2) node[right]{Homodyne detector} ;
    \draw (12,8) node[right]{Homodyne detector} ;
    \draw (6,2) node[above left]{PBS} ;
    \draw (10,2) node[above left]{TBS} ;
    \draw (10,8) node[below left]{TBS} ;
    \draw (7,5) node[right]{Photon detector} ;
    \draw (10,9) node[above]{$\rho_{da}$} ;
    \draw (6,8) node[below left]{BS} ;
    \draw (0,8) node[above]{Squeezed vacuum} ;
  \end{tikzpicture}
  \caption{A scheme of the set-up, with the theoretical beam-splitters (TBS) which purpose are to take into account noise in the set up.}
  \label{schemeofsetupvacuum}
\end{figure}

The experimental setup we consider here uses optical fields at room 
temperature, so it is reasonable to take $\rho_{ca} = \rho_{da} =
\ket{0}\bra{0}$. Indeed, for optical frequencies, the average 
number of thermal photon at room temperature is $\moy{n} = \frac{1}{e^{\frac{h \nu}{k_B T}} - 1} \approx 10^{-54}$. We nonetheless also considered the case where the fluctuating ancillary fields $\rho_{da}$ and $\rho_{ca}$ are thermal fields at finite temperature instead of vacuum, as shown in Appendix \ref{Thermalnoise}. It does not change our results qualitatively. 

An important aspect of the noise model we considered is that it does not increase the dimension of the pure state. Indeed, $\rho_{\text{noise}}$ can be represented as a $4\times 4$ matrix like the original $\ket{\psi}\bra{\psi}$, albeit in a different basis. The complete expression of $\rho_{\text{noise}}$ after performing the partial trace is given in Appendix \ref{ConcurrenceAppendix}. It is a ``mixed hybrid entangled states", according to the classification of Kreis and Van Loock in their seminal work \cite{kreis_classifying_2012,kreis_characterizing_2012}. Consequently, its entanglement can be studied analogously to a DV only system: one can define a subspace dependent Pauli-like algebra involving observables with a continuous spectrum in order to define an easy-to-implement EW. 

In order to simplify the expression of the noisy state, it is convenient to 
write it in the following orthonormal basis
\begin{align}
  \{\ket{C^+(\sqrt{1-\eta}\alpha)}\ket{0},\ket{C^+(\sqrt{1-\eta}\alpha)}\ket{1}, \\
  \ket{C^-(\sqrt{1-\eta}\alpha)}\ket{0},\ket{C^-(\sqrt{1-\eta}\alpha)}\ket{1}\}
  \label{dampedcat}
\end{align}

In this basis, $\rho_{\text{noise}}$ takes the following simple form:
\begin{equation}
  \rho_{\text{noise}} = \begin{pmatrix}
    w & 0 & 0 & z  \\
    0 & x_1 & c & 0 \\
    0 & c & x_2 & 0 \\
    z & 0 & 0 & y 
    \label{densitynoisymatrix1}
  \end{pmatrix} ,
\end{equation}
where $w,z,x_1,f,x_2,z$ and $y$ are functions of $\eta$, $\eta_d$ and 
$\alpha$, that are given in Appendix \ref{ConcurrenceAppendix}.

Another interesting aspect of being able to express the noisy state as a $4 \times 4$ system is that the photon loss noise model can be formulated as a quantum channel in terms of Kraus operators. For such, we write $\mathcal{U}(\eta)$ the operator performing the change of basis from $\left\{\ket{C^{\pm}(\alpha)}\right\}$ to the noise dependent basis 
$\left\{\ket{C^{\pm}(\eta\alpha)}\right\}$, for the continuous part.
Then the state $\rho_{\text{noise}}$ given by Eq.~\eqref{densitynoisymatrix1} 
can be obtained from the ideal state $\ket{\psi}\bra{\psi}$, with the help of local Kraus operators $\mathcal{C}_i\mathcal{U}(\eta) \otimes 
\mathcal{D}_j (i,j = 1,2)$ as:
\begin{equation}
  \rho_{\text{noise}} = \sum_{i,j=1}^2 \mathcal{C}_i\mathcal{U}(\sqrt{1 \text{-}\eta}\alpha)
  \otimes 
  \mathcal{D}_j  \ket{\psi}\bra{\psi}
  [\mathcal{C}_i\mathcal{U}(\sqrt{1\text{-}\eta}\alpha)]^{\dagger} 
  \otimes 
  \mathcal{D}_j^{\dagger} 
  \label{IntroKrauss}
\end{equation}
where the operators ($ \mathcal{C}_i$) and ($\mathcal{D}_j$) are calculated 
in Appendix \ref{NaimarkAppendix}. For the discrete part we obtain:
\begin{equation}
  \mathcal{D}_1 = \pmat{1}{0}{0}{\sqrt{1-\eta_d}}, \quad
  \mathcal{D}_2 = \pmat{0}{\sqrt{\eta_d}}{0}{0} 
\end{equation}
which is an amplitude damping channel.
The Krauss operators for the continuous part can be written as
\begin{equation}
  \mathcal{C}_1 = \pmat{\cos{\alpha}}{0}{0}{\cos{\beta}} ,\quad 
  \mathcal{C}_2 = \pmat{0}{\sin{\beta}}{\sin{\alpha}}{0}
\end{equation}
with 
$$\alpha = \arccos{\frac{\sqrt{(1+\exp{(- 2 (1-\eta) \alpha^2)})(1+\exp{(- 2 
  \eta \alpha^2)})}}{\sqrt{2+2 \exp{(- 2 \alpha^2)}
}}}$$
and 
$$\beta = \arccos{\frac{\sqrt{(1-\exp{(- 2 (1-\eta) \alpha^2)})(1+\exp{(- 2 
\eta \alpha^2)})}}{\sqrt{2-2 \exp{(- 2 \alpha^2)}}}}.$$
When $\alpha = \beta$ we obtain a dephasing channel, whereas when $\beta = 
0$ we have an amplitude-damping channel~\cite{wolf2007quantum}, so for the 
continuous part, aside from the unitary transformation $\mathcal{U}$,
the quantum channel is a combination of these two channels. 

An alternative encoding of DV quantum information for the discrete part of our hybrid state would use the polarisation degrees of freedom instead of the vacuum and one photon Fock state. In this case, the noise model would change, and it would be reasonable to consider instead a depolarizing channel on the discrete side. We can show that even in this case, the density matrix has the same form as the one presented in Equation \eqref{densitynoisymatrix1}.

\subsection{Entanglement Characterization}
As we have noted previously, for a given value of $\eta$ and $\eta_d$, the 
state $\rho_{\text{noise}}$ can be described by a $4 \times 4$ density matrix in an orthonormal basis which depends on the noise parameter $\eta$. This means that we can 
consider it as an effective $4 \times 4$ DV-system and completely characterise its entanglement \cite{van2001entangled, wang2001bipartite, kreis_characterizing_2012}. To this end, we choose the concurrence $C$ of $\rho_{\text{noise}}$, which takes the following simple form:\cite{santos2006direct}: 
\begin{equation}
  C(\rho_{\text{noise}}) = \max (0, 2c - 2 \sqrt{wy} ),
  \label{Concurrence}
\end{equation}
(See Appendix \ref{ConcurrenceAppendix}). For a 2 qubit system, as it is the case here, it is positive if and only if the state is entangled. 

\begin{figure}
  \begin{center}
    \includegraphics[width=1.\linewidth]{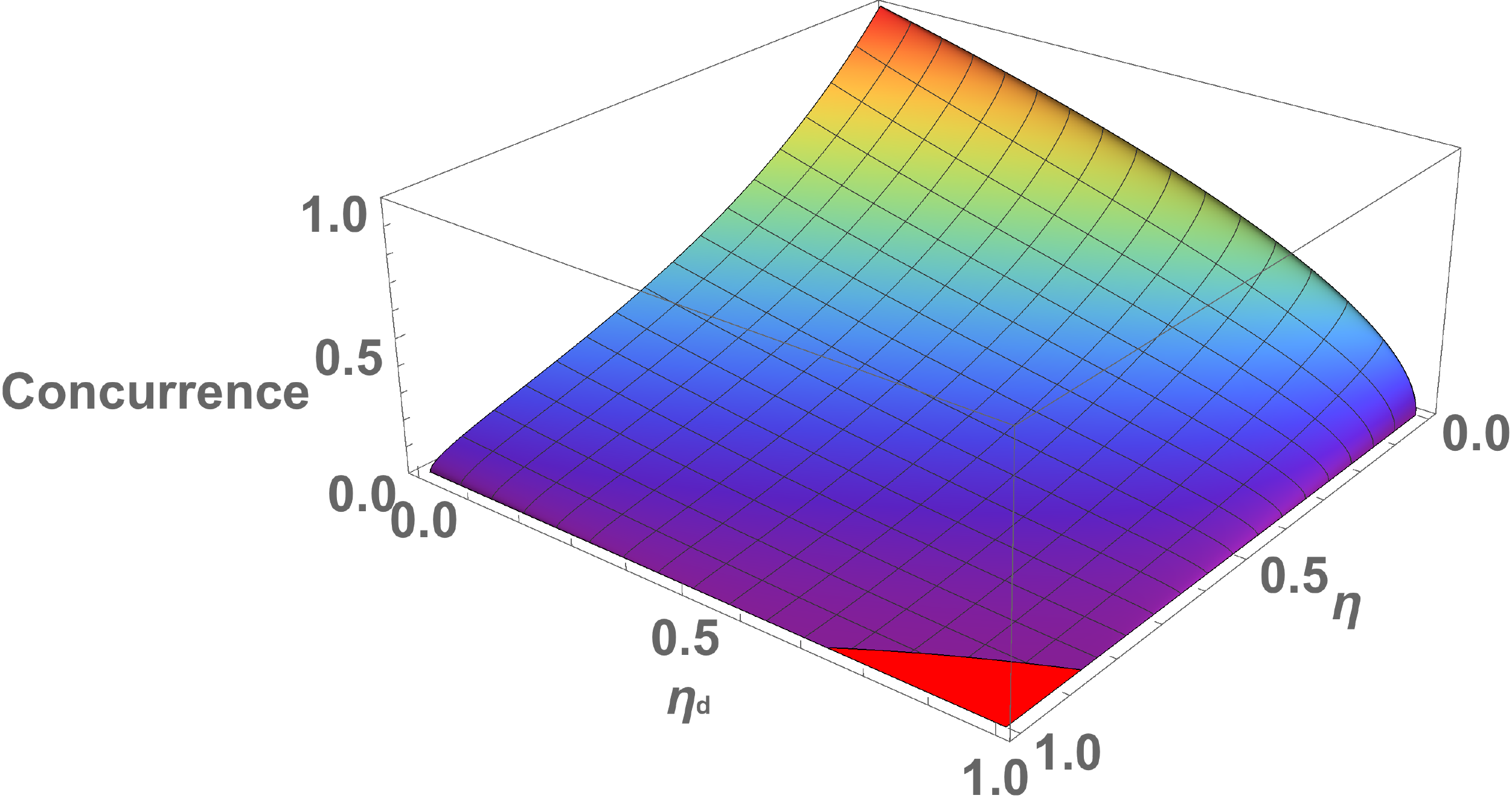}
    \caption{Concurrence $C(\rho_\text{noise})$ as a 
      function of noise 
      parameters $\eta$ and $\eta_d$ for an amplitude $\alpha = 1$. The 
      negative value are clipped, only the positive value indicating 
    entanglement of $\rho_\text{noise}$ are plotted (in rainbow colours).}
    \label{conccat1}
  \end{center}
\end{figure}

\begin{figure}
  \begin{center}
    \includegraphics[width=1.\linewidth]{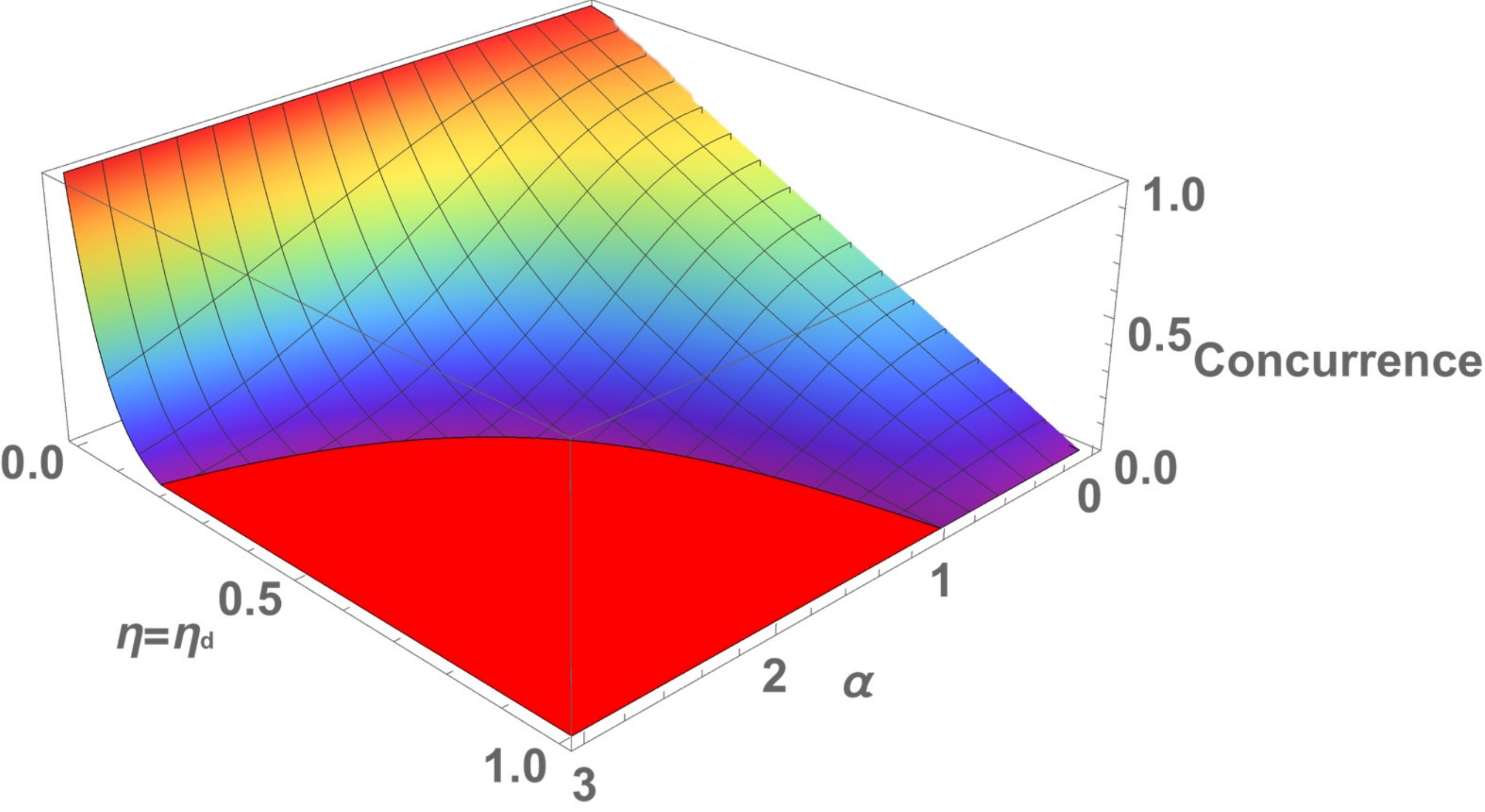}
    \caption{Concurrence $C(\rho_\text{noise})$ as a 
      function of the cat size 
      $\alpha$ and the noise $\eta=\eta_d$. The negative value are 
      clipped, only the positive value indicating entanglement of $\rho_\text{noise}$ are plotted (in rainbow colours).}
    \label{concUniqueT}
  \end{center}
\end{figure}

We show in Figures~\ref{conccat1} and~\ref{concUniqueT} the variation of 
the concurrence $C(\rho_{\text{noise}})$ as a function of noise 
parameters $\eta$ and $\eta_d$ and the amplitude $\alpha$. Figure~\ref{conccat1} shows that the 
concurrence is decreasing with respect to the amount of noise on 
each channel. With $\alpha = 1$, the state is separable only when the noise is very important ($\eta = \eta_d \geq 0.8$). Now, if we set $\eta = \eta_d$, we observe in 
Figure~\ref{concUniqueT} that the concurrence decreases with respect to 
$\alpha$ and $\eta$. Besides, the entanglement of the state becomes more and 
more sensitive to the noise, as the amplitude $\alpha$ increases. 

\section{ Entanglement witness}
We now consider the entanglement witness $W = \frac{1}{2}\mathds{1} - 
\ket{\psi} \bra{\psi}$. $\Tr[W\sigma]$ is positive for all separable state since the Schmidt rank of $\ket{\psi}$ can not exceed 2. This is due to the fact that the Schmidt rank is bounded by the Hilbert space with the lowest dimension value: the one of the qubit. $W$ is well suited to detect 
the target state $\ket{\psi}$ since
$\bra{\psi}W\ket{\psi} = -\frac{1}{2}<0$.

\subsection{Noise robustness}

The relevance and usefulness of $W$ is related to its ability to detect entanglement for a large set of $\rho_{\text{noise}}$ states. When computing $\Tr[W\rho_{\text{noise}}]$, we obtain
\begin{equation}
  \Tr[W\rho_{\text{noise}}] = \omega + y - 2c.
\end{equation}

\begin{figure}
  \begin{center}
    \includegraphics[width=1.\linewidth]{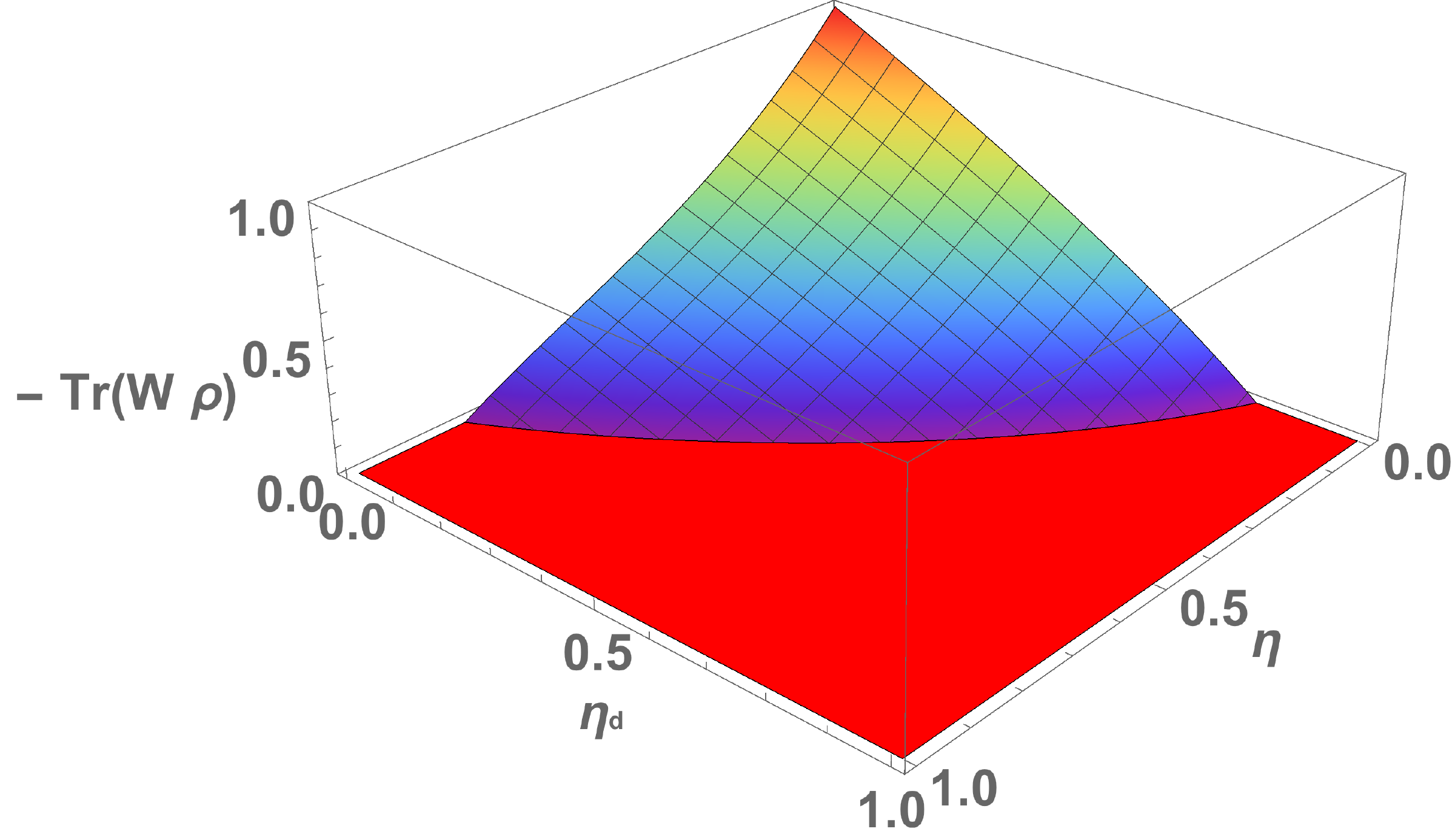}
    \caption{$- \Tr[W\rho_{\text{noise}}]$ as a function of the noise parameters 
      $\eta$ and $\eta_d$ for a cat size $\alpha=1$. The 
      negative value are clipped, only the positive value indicating 
    entanglement of $\rho_\text{noise}$ are plotted (in rainbow colours).}
    \label{witnesaprox}
  \end{center}
\end{figure}

\begin{figure}
  \begin{center}
    \includegraphics[width=1.\linewidth]{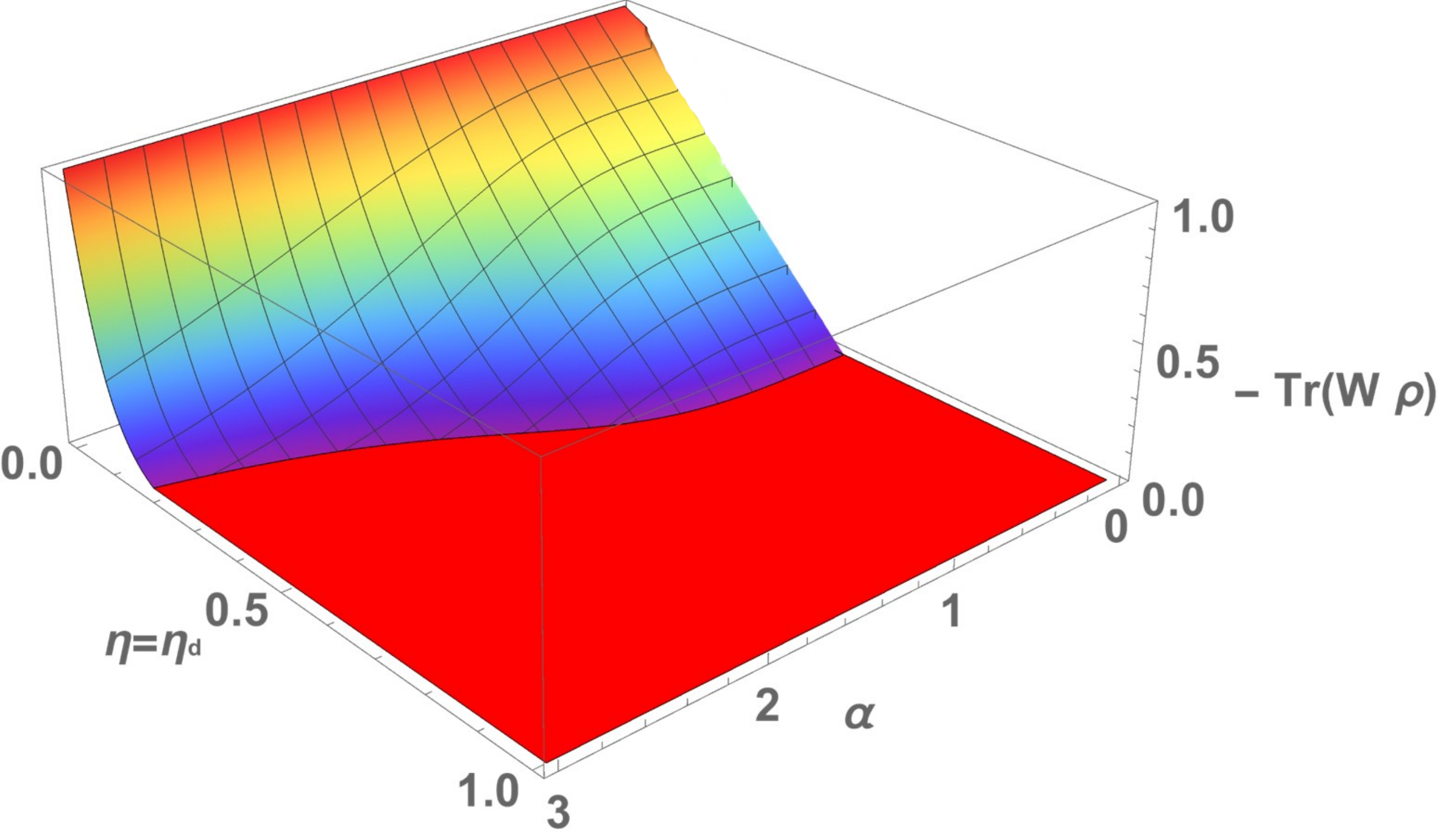}
    \caption{$-\Tr[W\rho_{\text{noise}}]$ as a function of the cat size $\alpha$ and the noise $\eta=\eta_d$. The negative value are clipped, only the positive value indicating entanglement of $\rho_\text{noise}$ are plotted(in rainbow colours).}
    \label{witnesaproxalpha}
  \end{center}
\end{figure}

We show in Figures~\ref{witnesaprox} and~\ref{witnesaproxalpha} the variation of 
$-\Tr[W\rho_{\text{noise}}]$ (we changed the sign to compare it more easily to the concurrence) as a function of noise 
parameters $\eta$ and $\eta_d$ and the amplitude $\alpha$ of the cat state.
Figure~\ref{witnesaprox} shows that $W$ detects entanglement even when both $\eta = \eta_d$ are equal to $0.5$ for $\alpha=1$. Since state-of-the-art optical set-ups can provide states with less than 20 \% of noise on each channel \cite{PhysRevLett.121.170403}, we consider that the robustness is satisfying. We can now discuss the witness implementation. By comparing Figure~\ref{witnesaproxalpha} with Figure~\ref{concUniqueT}, we see that for increasing $\alpha$, the region of non-detected entangled states in the form of (\ref{eq:catStateDef}) decreases: the witness tends more and more to become a necessary condition, \textit{i.e} $\Tr[W\rho_{\text{noise}}] \leq 0 \sim C(\rho_{\text{noise}})$.

\subsection{Experimental Implementation}

Measuring $W$ involves defining local projectors characterising $\ket{\psi}\bra{\psi}$ both on its discrete and its continuous parts. For the discrete part, we can safely consider the Pauli matrices 
$\sigma_z = \ket{0}\bra{0} - \ket{1}\bra{1}$, $\sigma_x = \ket{0}\bra{0} + 
\ket{1}\bra{1}$ and $\sigma_y = \frac{1}{2i}[\sigma_z,\sigma_x]$. For the 
continuous part, we can define analogous observables with a continuous spectrum, \textit{i.e} with the same matrix but in the \{$\ket{C^-(1-\eta)\alpha)}, \ket{C^+(1-\eta)\alpha)}$ \} basis. Specifically, 
\begin{align}
X_C &= \ket{C^-(1-\eta)\alpha)} \bra{C^+(1-\eta)\alpha)} \nonumber \\
&+ \ket{C^+(1-\eta)\alpha)} \bra{C^-(1-\eta)\alpha)},\\
Z_C &= \ket{C^-(1-\eta)\alpha)} \bra{C^+(1-\eta)\alpha)} \nonumber \\
&+ \ket{C^+(1-\eta)\alpha)} \bra{C^-(1-\eta)\alpha)}, \\
Y_C &= \frac{1}{2i}[Z_C,X_C].
\end{align}
This yields:
\begin{equation}
  4 \ket{\psi} \bra{\psi} = \left[ \mathds{1} + \sigma_x \otimes X_C - 
  \sigma_y \otimes Y_C + \sigma_z \otimes Z_C \right]
  \label{witnessobservables}
\end{equation}
Observables $X_C$, $Y_C$ and $Z_C$ are non Gaussian and can not be experimentally measured in a straightforward way. In order to propose an easy way to measure the witness, we can replace them by observables that reproduce a Pauli algebra in the specific subspace of interest, that of states given in Eq. (\ref{eq:catStateDef}). In such a subspace, we can replace:
\begin{equation}
  X_C \longrightarrow \frac{a+a^{\dagger}}{n_x},\quad Y_C \longrightarrow 
  \frac{i(a-a^{\dagger})}{n_y}
  \label{paulialgebrax}
\end{equation}
\begin{equation}
  Z_C \longrightarrow \lambda_z a^{\dagger}a + \mu_z
\end{equation}
where $n_x, n_y, \mu_z, \lambda_z$ are normalisation factors
depending weakly on the parameters $\alpha, \eta, \eta_d$ of the experiment. Such observables correspond to homodyne measurements at fixed angles. Hence, we define the new operator
\begin{equation}
  \tilde{W} = \mathds{1} - \frac{1}{2}\left[ \mathds{1} + \sigma_x \otimes 
    \frac{a+a^{\dagger}}{n_x} - \sigma_y\otimes
  \frac{i(a-a^{\dagger})}{n_y} \right]
  \label{finalwitness}
\end{equation}
which is now written in terms of observables which are currently measured 
in quantum optics experiments using homodyne detection ~\cite{van2011optical, morin_witnessing_2013, PhysRevLett.121.170403}. The term $\sigma_z \otimes Z_C$ has been discarded since it does not significantly change the value of $\Tr[W \rho_{\text{noise}}]$ and thus does not help to detect the entanglement of $\rho_{\text{noise}}$. As a matter of fact, it increases the difficulty to fulfil the condition $\Tr[ \tilde{W} \sigma] \geq 0 $ for all $\sigma$ separable. However, since $\tilde{W}$ is different from $W$, (their expectation values coincide only in the case of cat states) it is necessary to prove that $\tilde{W}$ is still an entanglement witness. Note that the values of $n_x$ and $n_y$ do not need anymore to be normalisation parameters: we can freely choose their values to optimise the witness.

We calculate in Appendix \ref{appendixwitness} an upper bound of the expectation value of $\tilde{W}$ for separable states. It depends on the number of photons in the continuous channel and the noise parameters $\eta$ and $\eta_d$. The proof involves approximating the Hilbert space of the continuous part of the hybrid state by a finite dimensional Hilbert space spanned by the Fock states $\{\ket{n}; \widehat{N} \ket{n} = n\ket{n} \text{ and } n\leq N\}$ where $\widehat{N}$ is the photon number operator. The value of the considered cut-off $N$ must of course increase when the cat size $\alpha$ increases, but this will have an impact on the ability of the witness to detect entanglement. Therefore, a balance must be found between the parameters $\alpha , \eta, \eta_d$ in order to detect the entanglement of $\rho_{\text{noise}}$. 

The detection of entanglement can now be carried out according to the following procedures: we choose a cut-off $N$, compute $n_x$ and $n_y$ such that no separable states within the sub-Hilbert space can violate the upper bound of the witness, and consider that the states we produce are in this subspace. This method is easy to test experimentally but over-evaluates the upper bound for separable states, as detailed in Appendix \ref{appendixwitness}, and necessitates the assumption that the states produced experimentally have no components on the Fock states $\ket{n}$ for $n \geq N$.

We propose a second method which requires additional measurements but does not necessitate to make this assumption, and that is more accurate with respect to the upper bound of the separable states. We explain it briefly here and more precisely in Appendix \ref{Control}. We use the method described in \cite{qi2020characterizing} to estimate the photon number distribution of the experimental states on the continuous channel. Using two conjugate homodyne detectors on this channel, we are able to measure simultaneously two orthogonal quadratures of the electromagnetic field. The sum of the square of these two output approximates sufficiently well the photon number operator $\widehat{N}$ to obtain the photon number distribution with a very good precision. Thanks to this knowledge, we are able to determine precisely the cut-off $N$ of the continuous channel without assumptions \textit{a priori}, and to compute an upper bound on the separable states more precise than the one obtained by Method 1. Finally, we also give in Appendix \ref{Control}, for experimental purposes, an alternative protocol for Method 2 which necessitates only one homodyne detector for the continuous channel but at the expense of the accuracy in the photon number distribution estimation. 


We summarise the two methods in the following table:

\begin{table}[h!]
  \centering
  \begin{adjustbox}{max width=250 pt}
    \begin{tabular}{|l|c|c|c|}
      \hline 
      \hspace{4cm} & Method 1 & \multicolumn{2}{c|}{Method 2}  \\
      \cline{3-4}
      \hline
      Assumption on the dimension & Yes & \multicolumn{2}{c|}{No}  \\
      \cline{3-4}
      \hline
      Evaluation of the photon statistic & No & \multicolumn{2}{c|}{Yes} \\
      \cline{3-4}
      \hline
      Number of homodyne detectors &  1 &  1 & 2 \\
      \hline
      Robustness to noise & Standard & Increased & Optimal\\
      \hline
    \end{tabular}
  \end{adjustbox}
  \label{WitnessComparison}
\end{table}

\begin{figure}
  \begin{center}
    \scalebox{.7}{\input{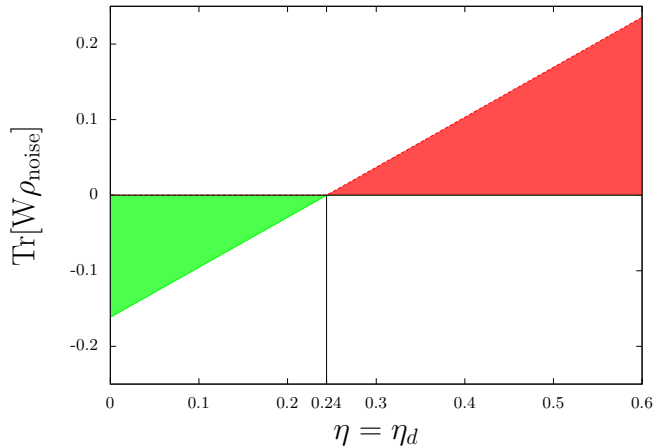}}
    \caption{$\Tr[\tilde{W}\rho_{\text{noise}}]$ as a function of the noise 
    parameters $\eta = \eta_d$ (same on both modes), with $\alpha= 1$ and cut-off at $N=3$. Entanglement is detected in the green zone, undetected in the red zone. $\eta_c = 0.24$.}
    \label{witnestrunca}
  \end{center}
\end{figure}

\begin{figure}
  \begin{center}
    \scalebox{.7}{\input{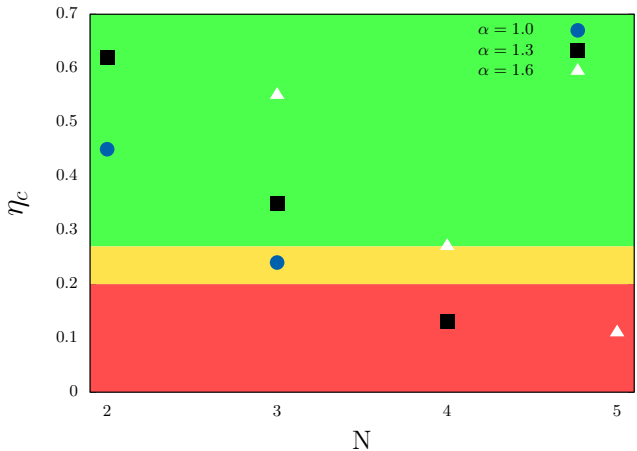}}
    \caption{Critical percentage of noise vs cut-off in the Fock space. The gold band corresponds to typical values of noise observed in state-of-the-art experiments \cite{lejeannic:tel-01665496}. Green zone shows a zone where the detection is experimentally easy. Red zone shows values of noise harder to obtain. $\eta_c = 0$ corresponds to an ideal case.}
    \label{NoisevsFock}
  \end{center}
\end{figure}

We illustrate Method 1 with two plots. Figure \ref{witnestrunca} shows the evolution of $\Tr[\tilde{W}\rho_{\text{noise}}]$ as a function of the noise parameters $\eta = \eta_d$, with $\alpha= 1$ and a cut-off at $N=3$. We see that the critical $\eta$ parameter, $\eta_c$ is equal to $22 \%$. We plot in Figure~\ref{NoisevsFock} $\eta_c$ against $N$ for $\alpha =1$, $\alpha =1.3$ and $\alpha =1.6$ to show the sensibility of $\eta_c$ to $N$ and $\alpha$. Method 2 is intended to be used on experiments. In order to test its relevance, we simulated experiments, like in part IV of \cite{qi2020characterizing}, with very good precision in the photon number distribution, that showed we could obtain $\eta_c \approx 20 \%$ for $\alpha = 1$, which is reasonable.

\section{Discussion}
In the present paper we considered the detection of a useful entangled state currently experimentally produced in quantum optics experiments. Our entanglement witness requires, to be evaluated, only the measurements of correlations between two Pauli matrices on the discrete side, and two quadratures of the field on the continuous side. Hence, contrary to the detection of a Wigner function or even of its negativity \cite{arkhipov2018negativity}, we do not need to measure displacement operators, nor do we need to use Photon Number Resolving (PNR) detectors \cite{laiho2009direct, sridhar2014direct}. The proposed witness can be measured using homodyne detectors in both sides, discrete and continuous. This  would only require to lock the phase of the local oscillator at two angles, to obtain two orthogonal quadratures $\hat{x}$ and $\hat{p}$, whereas in a full tomography the measurement of all possible orthogonal quadratures is required. 

We summarise the proposed measurement protocol as follows :

\begin{framed} 
  \begin{enumerate}
    \item Lock the phase of the local oscillators on the homodyne detectors to detect \^x
    \item Record data on both sides
    \item Compute correlations $\moy{\sigma_X \otimes 
      \frac{a+a^{\dagger}}{n_x(\alpha, \eta_X)}}_{\rho_{\text{exp}}}$
    \item Lock the phase of the local oscillators on the homodyne detectors to detect \^p
    \item Record data on both sides
    \item Compute correlations $\moy{\sigma_Y \otimes 
      \frac{a-a^{\dagger}}{n_y(\alpha, \eta_Y)}}_{\rho_{\text{exp}}}$
    \item If Method 2 is chosen, compute the bound on the separable states
    \item Compute the value of the Witness
  \end{enumerate}
  \label{Protocol}
\end{framed}

\section{Conclusion}

We have presented an implementable hybrid entanglement witness. that can be experimentally detected with only a few relatively easy to perform measurements. This was achieved, in a first step, by identifying observables with a continuous spectrum to Pauli matrices in a specific subspace. Such identification was possible thanks to the fact that noise, in the considered subspace, does not increase its dimension. In a second step, we replaced such observables by others, easier to measure, that coincide within the targeted subspace. We hope this work can help to understand better the subtle features of hybrid entanglement and, more generally, hybrid quantum protocols, both theoretically and experimentally. 

\section*{Acknowledgments}
We acknowledge fruitful discussions with T. Darras, J. Laurat, L. Garbe and N. Fabre. G.M. acknowledges support from the French Agence Nationale de la Recherche (ANR-17-CE30-0006).

\clearpage
\small{\bibliographystyle{unsrt}
\bibliography{Report2}\vspace{0.75in}}

\begin{thebibliography}{53}
\expandafter\ifx\csname natexlab\endcsname\relax\def\natexlab#1{#1}\fi
\expandafter\ifx\csname bibnamefont\endcsname\relax
  \def\bibnamefont#1{#1}\fi
\expandafter\ifx\csname bibfnamefont\endcsname\relax
  \def\bibfnamefont#1{#1}\fi
\expandafter\ifx\csname citenamefont\endcsname\relax
  \def\citenamefont#1{#1}\fi
\expandafter\ifx\csname url\endcsname\relax
  \def\url#1{\texttt{#1}}\fi
\expandafter\ifx\csname urlprefix\endcsname\relax\def\urlprefix{URL }\fi
\providecommand{\bibinfo}[2]{#2}
\providecommand{\eprint}[2][]{\url{#2}}

\bibitem[{\citenamefont{Van~Loock}(2011)}]{van2011optical}
\bibinfo{author}{\bibfnamefont{P.}~\bibnamefont{Van~Loock}},
  \bibinfo{journal}{Laser \& Photonics Reviews} \textbf{\bibinfo{volume}{5}},
  \bibinfo{pages}{167} (\bibinfo{year}{2011}).

\bibitem[{\citenamefont{Qi and Hou}(2016)}]{qi_nonlinear_2016}
\bibinfo{author}{\bibfnamefont{X.}~\bibnamefont{Qi}} \bibnamefont{and}
  \bibinfo{author}{\bibfnamefont{J.}~\bibnamefont{Hou}},
  \bibinfo{journal}{Quantum Information Processing}
  \textbf{\bibinfo{volume}{15}}, \bibinfo{pages}{741} (\bibinfo{year}{2016}),
  ISSN \bibinfo{issn}{1573-1332},
  \urlprefix\url{https://doi.org/10.1007/s11128-015-1156-0}.

\bibitem[{\citenamefont{Takeda and Furusawa}(2019)}]{takeda2019toward}
\bibinfo{author}{\bibfnamefont{S.}~\bibnamefont{Takeda}} \bibnamefont{and}
  \bibinfo{author}{\bibfnamefont{A.}~\bibnamefont{Furusawa}},
  \bibinfo{journal}{APL Photonics} \textbf{\bibinfo{volume}{4}},
  \bibinfo{pages}{060902} (\bibinfo{year}{2019}).



\bibitem[{\citenamefont{Takeda et~al.}(2013)\citenamefont{Takeda, Mizuta, Fuwa,
  van Loock, and Furusawa}}]{takeda_deterministic_2013}
\bibinfo{author}{\bibfnamefont{S.}~\bibnamefont{Takeda}},
  \bibinfo{author}{\bibfnamefont{T.}~\bibnamefont{Mizuta}},
  \bibinfo{author}{\bibfnamefont{M.}~\bibnamefont{Fuwa}},
  \bibinfo{author}{\bibfnamefont{P.}~\bibnamefont{van Loock}},
  \bibnamefont{and} \bibinfo{author}{\bibfnamefont{A.}~\bibnamefont{Furusawa}},
  \bibinfo{journal}{Nature} \textbf{\bibinfo{volume}{500}},
  \bibinfo{pages}{315} (\bibinfo{year}{2013}), ISSN \bibinfo{issn}{0028-0836,
  1476-4687}, \urlprefix\url{http://www.nature.com/articles/nature12366}.

\bibitem[{\citenamefont{Lee and Jeong}(2013)}]{lee_near-deterministic_2013}
\bibinfo{author}{\bibfnamefont{S.-W.} \bibnamefont{Lee}} \bibnamefont{and}
  \bibinfo{author}{\bibfnamefont{H.}~\bibnamefont{Jeong}},
  \bibinfo{journal}{Physical Review A} \textbf{\bibinfo{volume}{87}}
  (\bibinfo{year}{2013}), ISSN \bibinfo{issn}{1050-2947, 1094-1622},
  \urlprefix\url{https://link.aps.org/doi/10.1103/PhysRevA.87.022326}.

\bibitem[{\citenamefont{Lie and Jeong}(2019)}]{lie2019limitations}
\bibinfo{author}{\bibfnamefont{S.~H.} \bibnamefont{Lie}} \bibnamefont{and}
  \bibinfo{author}{\bibfnamefont{H.}~\bibnamefont{Jeong}},
  \bibinfo{journal}{Photonics Research} \textbf{\bibinfo{volume}{7}},
  \bibinfo{pages}{A7} (\bibinfo{year}{2019}).

\bibitem[{\citenamefont{Brask et~al.}(2012)\citenamefont{Brask, Brunner,
  Cavalcanti, and Leverrier}}]{brask2012bell}
\bibinfo{author}{\bibfnamefont{J.~B.} \bibnamefont{Brask}},
  \bibinfo{author}{\bibfnamefont{N.}~\bibnamefont{Brunner}},
  \bibinfo{author}{\bibfnamefont{D.}~\bibnamefont{Cavalcanti}},
  \bibnamefont{and}
  \bibinfo{author}{\bibfnamefont{A.}~\bibnamefont{Leverrier}},
  \bibinfo{journal}{Physical Review A} \textbf{\bibinfo{volume}{85}},
  \bibinfo{pages}{042116} (\bibinfo{year}{2012}).

\bibitem[{\citenamefont{Quintino et~al.}(2012)\citenamefont{Quintino,
  Ara{\'u}jo, Cavalcanti, Santos, and Cunha}}]{quintino2012maximal}
\bibinfo{author}{\bibfnamefont{M.~T.} \bibnamefont{Quintino}},
  \bibinfo{author}{\bibfnamefont{M.}~\bibnamefont{Ara{\'u}jo}},
  \bibinfo{author}{\bibfnamefont{D.}~\bibnamefont{Cavalcanti}},
  \bibinfo{author}{\bibfnamefont{M.~F.} \bibnamefont{Santos}},
  \bibnamefont{and} \bibinfo{author}{\bibfnamefont{M.~T.} \bibnamefont{Cunha}},
  \bibinfo{journal}{Journal of Physics A: Mathematical and Theoretical}
  \textbf{\bibinfo{volume}{45}}, \bibinfo{pages}{215308}
  (\bibinfo{year}{2012}).

\bibitem[{\citenamefont{Kwon and Jeong}(2013)}]{kwon2013violation}
\bibinfo{author}{\bibfnamefont{H.}~\bibnamefont{Kwon}} \bibnamefont{and}
  \bibinfo{author}{\bibfnamefont{H.}~\bibnamefont{Jeong}},
  \bibinfo{journal}{Physical Review A} \textbf{\bibinfo{volume}{88}},
  \bibinfo{pages}{052127} (\bibinfo{year}{2013}).

\bibitem[{\citenamefont{Töppel and Stobi{\'{n}}ska}(2015)}]{T_ppel_2015}
\bibinfo{author}{\bibfnamefont{F.}~\bibnamefont{Töppel}} \bibnamefont{and}
  \bibinfo{author}{\bibfnamefont{M.}~\bibnamefont{Stobi{\'{n}}ska}},
  \bibinfo{journal}{Journal of Physics A: Mathematical and Theoretical}
  \textbf{\bibinfo{volume}{48}}, \bibinfo{pages}{075306}
  (\bibinfo{year}{2015}),
  \urlprefix\url{https://doi.org/10.1088%2F1751-8113%2F48%2F7%2F075306}.

\bibitem[{\citenamefont{Brask et~al.}(2010)\citenamefont{Brask, Rigas, Polzik,
  Andersen, and S{\o}rensen}}]{brask2010hybrid}
\bibinfo{author}{\bibfnamefont{J.~B.} \bibnamefont{Brask}},
  \bibinfo{author}{\bibfnamefont{I.}~\bibnamefont{Rigas}},
  \bibinfo{author}{\bibfnamefont{E.~S.} \bibnamefont{Polzik}},
  \bibinfo{author}{\bibfnamefont{U.~L.} \bibnamefont{Andersen}},
  \bibnamefont{and} \bibinfo{author}{\bibfnamefont{A.~S.}
  \bibnamefont{S{\o}rensen}}, \bibinfo{journal}{Physical review letters}
  \textbf{\bibinfo{volume}{105}}, \bibinfo{pages}{160501}
  (\bibinfo{year}{2010}).

\bibitem[{\citenamefont{Lim et~al.}(2016)\citenamefont{Lim, Joo, Spiller, and
  Jeong}}]{lim_loss-resilient_2016}
\bibinfo{author}{\bibfnamefont{Y.}~\bibnamefont{Lim}},
  \bibinfo{author}{\bibfnamefont{J.}~\bibnamefont{Joo}},
  \bibinfo{author}{\bibfnamefont{T.~P.} \bibnamefont{Spiller}},
  \bibnamefont{and} \bibinfo{author}{\bibfnamefont{H.}~\bibnamefont{Jeong}},
  \bibinfo{journal}{Physical Review A} \textbf{\bibinfo{volume}{94}}
  (\bibinfo{year}{2016}), ISSN \bibinfo{issn}{2469-9926, 2469-9934},
  \urlprefix\url{https://link.aps.org/doi/10.1103/PhysRevA.94.062337}.

\bibitem[{\citenamefont{Andersen et~al.}(2015)\citenamefont{Andersen,
  Neergaard-Nielsen, van Loock, and Furusawa}}]{andersen_hybrid_2015}
\bibinfo{author}{\bibfnamefont{U.~L.} \bibnamefont{Andersen}},
  \bibinfo{author}{\bibfnamefont{J.~S.} \bibnamefont{Neergaard-Nielsen}},
  \bibinfo{author}{\bibfnamefont{P.}~\bibnamefont{van Loock}},
  \bibnamefont{and} \bibinfo{author}{\bibfnamefont{A.}~\bibnamefont{Furusawa}},
  \bibinfo{journal}{Nature Physics} \textbf{\bibinfo{volume}{11}},
  \bibinfo{pages}{713} (\bibinfo{year}{2015}), ISSN \bibinfo{issn}{1745-2473,
  1745-2481}, \bibinfo{note}{arXiv: 1409.3719},
  \urlprefix\url{http://arxiv.org/abs/1409.3719}.

\bibitem[{\citenamefont{Sychev et~al.}(2015)\citenamefont{Sychev, Ulanov,
 Tiunov, Pushkina, Kuzhamuratov, Novikov and Lvovsky }}]{sychev2018}
\bibinfo{author}{\bibfnamefont{D.~V.} \bibnamefont{Sychev}},
  \bibinfo{author}{\bibfnamefont{A.~E.} \bibnamefont{Ulanov}},
  \bibinfo{author}{\bibfnamefont{E.~S.}~\bibnamefont{Tiunov}},
  \bibinfo{author}{\bibfnamefont{A.~A.} \bibnamefont{Pushkina}},
  \bibinfo{author}{\bibfnamefont{A.} \bibnamefont{Kuzhamuratov}},
  \bibinfo{author}{\bibfnamefont{V.}~\bibnamefont{Novikov}},
  \bibnamefont{and} \bibinfo{author}{\bibfnamefont{A.~I.}~\bibnamefont{Lvovsky}},
  \bibinfo{journal}{Nature communications} \textbf{\bibinfo{volume}{9}},
  \bibinfo{pages}{1--7} (\bibinfo{year}{2018}), ISSN \bibinfo{issn}{2041-1723},
  \bibinfo{note}{arXiv: 1712.10206},
  \urlprefix\url{http://arxiv.org/abs/1712.10206}.

\bibitem[{\citenamefont{Huang et~al.}(2019)\citenamefont{Huang, Le~Jeannic,
  Morin, Darras, Guccione, Cavaill\`es, and Laurat}}]{engineeringopticalhybrid}
\bibinfo{author}{\bibfnamefont{K.}~\bibnamefont{Huang}},
  \bibinfo{author}{\bibfnamefont{H.}~\bibnamefont{Le~Jeannic}},
  \bibinfo{author}{\bibfnamefont{O.}~\bibnamefont{Morin}},
  \bibinfo{author}{\bibfnamefont{T.}~\bibnamefont{Darras}},
  \bibinfo{author}{\bibfnamefont{G.}~\bibnamefont{Guccione}},
  \bibinfo{author}{\bibfnamefont{A.}~\bibnamefont{Cavaill\`es}},
  \bibnamefont{and} \bibinfo{author}{\bibfnamefont{J.}~\bibnamefont{Laurat}}
  (\bibinfo{year}{2019}).

\bibitem[{\citenamefont{Guccione Darras et~al.}(2020)\citenamefont{Guccione, Darras,
 Le Jeannic, Verma, Nam, Cavaill\`es, Laurat}}]{guccione2020}
\bibinfo{author}{\bibfnamefont{G.}~\bibnamefont{Guccione}},
  \bibinfo{author}{\bibfnamefont{T.}~\bibnamefont{Darras}},
  \bibinfo{author}{\bibfnamefont{H.}~\bibnamefont{Le Jeannic}},
  \bibinfo{author}{\bibfnamefont{V. B.}~\bibnamefont{Verma}},
  \bibinfo{author}{\bibfnamefont{S. W.}~\bibnamefont{Nam}},
  \bibinfo{author}{\bibfnamefont{A.}~\bibnamefont{Cavaill\`es}}, \bibnamefont{and}
  \bibinfo{author}{\bibfnamefont{J.}~\bibnamefont{Laurat}},
  \bibinfo{journal}{arXiv preprint arXiv:2003.11041}  (\bibinfo{year}{2020}).

\bibitem[{\citenamefont{Cavaill\`es et~al.}(2018)\citenamefont{Cavaill\`es,
  Le~Jeannic, Raskop, Guccione, Markham, Diamanti, Shaw, Verma, Nam, and
  Laurat}}]{PhysRevLett.121.170403}
\bibinfo{author}{\bibfnamefont{A.}~\bibnamefont{Cavaill\`es}},
  \bibinfo{author}{\bibfnamefont{H.}~\bibnamefont{Le~Jeannic}},
  \bibinfo{author}{\bibfnamefont{J.}~\bibnamefont{Raskop}},
  \bibinfo{author}{\bibfnamefont{G.}~\bibnamefont{Guccione}},
  \bibinfo{author}{\bibfnamefont{D.}~\bibnamefont{Markham}},
  \bibinfo{author}{\bibfnamefont{E.}~\bibnamefont{Diamanti}},
  \bibinfo{author}{\bibfnamefont{M.~D.} \bibnamefont{Shaw}},
  \bibinfo{author}{\bibfnamefont{V.~B.} \bibnamefont{Verma}},
  \bibinfo{author}{\bibfnamefont{S.~W.} \bibnamefont{Nam}}, \bibnamefont{and}
  \bibinfo{author}{\bibfnamefont{J.}~\bibnamefont{Laurat}},
  \bibinfo{journal}{Phys. Rev. Lett.} \textbf{\bibinfo{volume}{121}},
  \bibinfo{pages}{170403} (\bibinfo{year}{2018}),
  \urlprefix\url{https://link.aps.org/doi/10.1103/PhysRevLett.121.170403}.

\bibitem[{\citenamefont{Bergmann and van
  Loock}(2019)}]{PhysRevA.99.032349Repeaters}
\bibinfo{author}{\bibfnamefont{M.}~\bibnamefont{Bergmann}} \bibnamefont{and}
  \bibinfo{author}{\bibfnamefont{P.}~\bibnamefont{van Loock}},
  \bibinfo{journal}{Phys. Rev. A} \textbf{\bibinfo{volume}{99}},
  \bibinfo{pages}{032349} (\bibinfo{year}{2019}),
  \urlprefix\url{https://link.aps.org/doi/10.1103/PhysRevA.99.032349}.

\bibitem[{\citenamefont{Kwon and Jeong}(2015)}]{kwon_generation_2015}
\bibinfo{author}{\bibfnamefont{H.}~\bibnamefont{Kwon}} \bibnamefont{and}
  \bibinfo{author}{\bibfnamefont{H.}~\bibnamefont{Jeong}},
  \bibinfo{journal}{Physical Review A} \textbf{\bibinfo{volume}{91}}
  (\bibinfo{year}{2015}), ISSN \bibinfo{issn}{1050-2947, 1094-1622},
  \bibinfo{note}{arXiv: 1410.6823},
  \urlprefix\url{http://arxiv.org/abs/1410.6823}.

\bibitem[{\citenamefont{Chitambar and Gour}(2019)}]{RevModPhys.91.025001}
\bibinfo{author}{\bibfnamefont{E.}~\bibnamefont{Chitambar}} \bibnamefont{and}
  \bibinfo{author}{\bibfnamefont{G.}~\bibnamefont{Gour}},
  \bibinfo{journal}{Rev. Mod. Phys.} \textbf{\bibinfo{volume}{91}},
  \bibinfo{pages}{025001} (\bibinfo{year}{2019}),
  \urlprefix\url{https://link.aps.org/doi/10.1103/RevModPhys.91.025001}.

\bibitem[{\citenamefont{Horodecki et~al.}(2009)\citenamefont{Horodecki,
  Horodecki, Horodecki, and Horodecki}}]{horodecki2009quantum}
\bibinfo{author}{\bibfnamefont{R.}~\bibnamefont{Horodecki}},
  \bibinfo{author}{\bibfnamefont{P.}~\bibnamefont{Horodecki}},
  \bibinfo{author}{\bibfnamefont{M.}~\bibnamefont{Horodecki}},
  \bibnamefont{and}
  \bibinfo{author}{\bibfnamefont{K.}~\bibnamefont{Horodecki}},
  \bibinfo{journal}{Reviews of modern physics} \textbf{\bibinfo{volume}{81}},
  \bibinfo{pages}{865} (\bibinfo{year}{2009}).

\bibitem[{\citenamefont{Horodecki et~al.}(1996)\citenamefont{Horodecki,
  Horodecki, and Horodecki}}]{horodecki_separability_1996}
\bibinfo{author}{\bibfnamefont{M.}~\bibnamefont{Horodecki}},
  \bibinfo{author}{\bibfnamefont{P.}~\bibnamefont{Horodecki}},
  \bibnamefont{and}
  \bibinfo{author}{\bibfnamefont{R.}~\bibnamefont{Horodecki}},
  \bibinfo{journal}{Physics Letters A} \textbf{\bibinfo{volume}{223}},
  \bibinfo{pages}{1} (\bibinfo{year}{1996}), ISSN \bibinfo{issn}{03759601},
  \bibinfo{note}{arXiv: quant-ph/9605038},
  \urlprefix\url{http://arxiv.org/abs/quant-ph/9605038}.

\bibitem[{\citenamefont{Chruściński and
  Sarbicki}(2014)}]{chruscinski_entanglement_2014}
\bibinfo{author}{\bibfnamefont{D.}~\bibnamefont{Chruściński}}
  \bibnamefont{and} \bibinfo{author}{\bibfnamefont{G.}~\bibnamefont{Sarbicki}},
  \bibinfo{journal}{Journal of Physics A: Mathematical and Theoretical}
  \textbf{\bibinfo{volume}{47}}, \bibinfo{pages}{483001}
  (\bibinfo{year}{2014}), ISSN \bibinfo{issn}{1751-8113, 1751-8121},
  \bibinfo{note}{arXiv: 1402.2413},
  \urlprefix\url{http://arxiv.org/abs/1402.2413}.

\bibitem[{\citenamefont{Hyllus et~al.}(2005)\citenamefont{Hyllus, G\"uhne,
  Bruss, and Lewenstein}}]{hyllus_relations_2005}
\bibinfo{author}{\bibfnamefont{P.}~\bibnamefont{Hyllus}},
  \bibinfo{author}{\bibfnamefont{O.}~\bibnamefont{G\"uhne}},
  \bibinfo{author}{\bibfnamefont{D.}~\bibnamefont{Bruss}}, \bibnamefont{and}
  \bibinfo{author}{\bibfnamefont{M.}~\bibnamefont{Lewenstein}},
  \bibinfo{journal}{Physical Review A} \textbf{\bibinfo{volume}{72}}
  (\bibinfo{year}{2005}), ISSN \bibinfo{issn}{1050-2947, 1094-1622},
  \bibinfo{note}{arXiv: quant-ph/0504079},
  \urlprefix\url{http://arxiv.org/abs/quant-ph/0504079}.

\bibitem[{\citenamefont{G\"uhne and Toth}(2009)}]{guhne_entanglement_2009}
\bibinfo{author}{\bibfnamefont{O.}~\bibnamefont{G\"uhne}} \bibnamefont{and}
  \bibinfo{author}{\bibfnamefont{G.}~\bibnamefont{Toth}},
  \bibinfo{journal}{Physics Reports} \textbf{\bibinfo{volume}{474}},
  \bibinfo{pages}{1} (\bibinfo{year}{2009}), ISSN \bibinfo{issn}{03701573},
  \bibinfo{note}{arXiv: 0811.2803},
  \urlprefix\url{http://arxiv.org/abs/0811.2803}.

\bibitem[{\citenamefont{Sperling and Vogel}(2009)}]{sperling_necessary_2009}
\bibinfo{author}{\bibfnamefont{J.}~\bibnamefont{Sperling}} \bibnamefont{and}
  \bibinfo{author}{\bibfnamefont{W.}~\bibnamefont{Vogel}},
  \bibinfo{journal}{Physical Review A} \textbf{\bibinfo{volume}{79}}
  (\bibinfo{year}{2009}), ISSN \bibinfo{issn}{1050-2947, 1094-1622},
  \urlprefix\url{https://link.aps.org/doi/10.1103/PhysRevA.79.022318}.

\bibitem[{\citenamefont{Kreis and van Loock}(2012)}]{kreis_classifying_2012}
\bibinfo{author}{\bibfnamefont{K.}~\bibnamefont{Kreis}} \bibnamefont{and}
  \bibinfo{author}{\bibfnamefont{P.}~\bibnamefont{van Loock}},
  \bibinfo{journal}{Physical Review A} \textbf{\bibinfo{volume}{85}}
  (\bibinfo{year}{2012}), ISSN \bibinfo{issn}{1050-2947, 1094-1622},
  \bibinfo{note}{arXiv: 1111.0478},
  \urlprefix\url{http://arxiv.org/abs/1111.0478}.

\bibitem[{\citenamefont{Peres}(1996)}]{peres_separability_1996}
\bibinfo{author}{\bibfnamefont{A.}~\bibnamefont{Peres}},
  \bibinfo{journal}{Physical Review Letters} \textbf{\bibinfo{volume}{77}},
  \bibinfo{pages}{1413} (\bibinfo{year}{1996}), ISSN \bibinfo{issn}{0031-9007,
  1079-7114},
  \urlprefix\url{https://link.aps.org/doi/10.1103/PhysRevLett.77.1413}.

\bibitem[{\citenamefont{Simon}(2000{\natexlab{a}})}]{simon_peres-horodecki_2000}
\bibinfo{author}{\bibfnamefont{R.}~\bibnamefont{Simon}},
  \bibinfo{journal}{Physical Review Letters} \textbf{\bibinfo{volume}{84}},
  \bibinfo{pages}{2726} (\bibinfo{year}{2000}{\natexlab{a}}), ISSN
  \bibinfo{issn}{0031-9007, 1079-7114}, \bibinfo{note}{arXiv:
  quant-ph/9909044}, \urlprefix\url{http://arxiv.org/abs/quant-ph/9909044}.

\bibitem[{\citenamefont{Arkhipov et~al.}(2018)\citenamefont{Arkhipov,
  Barasi{\'n}ski, and Svozil{\'\i}k}}]{arkhipov2018negativity}
\bibinfo{author}{\bibfnamefont{I.~I.} \bibnamefont{Arkhipov}},
  \bibinfo{author}{\bibfnamefont{A.}~\bibnamefont{Barasi{\'n}ski}},
  \bibnamefont{and}
  \bibinfo{author}{\bibfnamefont{J.}~\bibnamefont{Svozil{\'\i}k}},
  \bibinfo{journal}{Scientific reports} \textbf{\bibinfo{volume}{8}},
  \bibinfo{pages}{16955} (\bibinfo{year}{2018}).

\bibitem[{\citenamefont{Hou and Qi}(2010)}]{hou_constructing_2010}
\bibinfo{author}{\bibfnamefont{J.}~\bibnamefont{Hou}} \bibnamefont{and}
  \bibinfo{author}{\bibfnamefont{X.}~\bibnamefont{Qi}},
  \bibinfo{journal}{Physical Review A} \textbf{\bibinfo{volume}{81}}
  (\bibinfo{year}{2010}), ISSN \bibinfo{issn}{1050-2947, 1094-1622},
  \bibinfo{note}{arXiv: 1005.5530},
  \urlprefix\url{http://arxiv.org/abs/1005.5530}.

\bibitem[{\citenamefont{Guo et~al.}(2011)\citenamefont{Guo, Qi, and
  Hou}}]{guo_sufficient_2011}
\bibinfo{author}{\bibfnamefont{Y.}~\bibnamefont{Guo}},
  \bibinfo{author}{\bibfnamefont{X.}~\bibnamefont{Qi}}, \bibnamefont{and}
  \bibinfo{author}{\bibfnamefont{J.}~\bibnamefont{Hou}},
  \bibinfo{journal}{Chinese Science Bulletin} \textbf{\bibinfo{volume}{56}},
  \bibinfo{pages}{840} (\bibinfo{year}{2011}), ISSN \bibinfo{issn}{1001-6538,
  1861-9541},
  \urlprefix\url{http://link.springer.com/10.1007/s11434-010-4500-x}.

\bibitem[{\citenamefont{Miranowicz et~al.}(2009)\citenamefont{Miranowicz,
  Piani, Horodecki, and Horodecki}}]{miranowicz_inseparability_2009}
\bibinfo{author}{\bibfnamefont{A.}~\bibnamefont{Miranowicz}},
  \bibinfo{author}{\bibfnamefont{M.}~\bibnamefont{Piani}},
  \bibinfo{author}{\bibfnamefont{P.}~\bibnamefont{Horodecki}},
  \bibnamefont{and}
  \bibinfo{author}{\bibfnamefont{R.}~\bibnamefont{Horodecki}},
  \bibinfo{journal}{Physical Review A} \textbf{\bibinfo{volume}{80}}
  (\bibinfo{year}{2009}), ISSN \bibinfo{issn}{1050-2947, 1094-1622},
  \bibinfo{note}{arXiv: quant-ph/0605001},
  \urlprefix\url{http://arxiv.org/abs/quant-ph/0605001}.

\bibitem[{\citenamefont{Gittsovich et~al.}(2015)\citenamefont{Gittsovich,
  Moroder, Asadian, G\"uhne, and Rabl}}]{gittsovich_non-classicality_2015}
\bibinfo{author}{\bibfnamefont{O.}~\bibnamefont{Gittsovich}},
  \bibinfo{author}{\bibfnamefont{T.}~\bibnamefont{Moroder}},
  \bibinfo{author}{\bibfnamefont{A.}~\bibnamefont{Asadian}},
  \bibinfo{author}{\bibfnamefont{O.}~\bibnamefont{G\"uhne}}, \bibnamefont{and}
  \bibinfo{author}{\bibfnamefont{P.}~\bibnamefont{Rabl}},
  \bibinfo{journal}{Physical Review A} \textbf{\bibinfo{volume}{91}}
  (\bibinfo{year}{2015}), ISSN \bibinfo{issn}{1050-2947, 1094-1622},
  \bibinfo{note}{arXiv: 1412.2167},
  \urlprefix\url{http://arxiv.org/abs/1412.2167}.

\bibitem[{\citenamefont{Shchukin and Vogel}(2005)}]{PhysRevLett.95.230502}
\bibinfo{author}{\bibfnamefont{E.}~\bibnamefont{Shchukin}} \bibnamefont{and}
  \bibinfo{author}{\bibfnamefont{W.}~\bibnamefont{Vogel}},
  \bibinfo{journal}{Phys. Rev. Lett.} \textbf{\bibinfo{volume}{95}},
  \bibinfo{pages}{230502} (\bibinfo{year}{2005}),
  \urlprefix\url{https://link.aps.org/doi/10.1103/PhysRevLett.95.230502}.

\bibitem[{\citenamefont{Simon}(2000{\natexlab{b}})}]{PhysRevLett.84.2726}
\bibinfo{author}{\bibfnamefont{R.}~\bibnamefont{Simon}},
  \bibinfo{journal}{Phys. Rev. Lett.} \textbf{\bibinfo{volume}{84}},
  \bibinfo{pages}{2726} (\bibinfo{year}{2000}{\natexlab{b}}),
  \urlprefix\url{https://link.aps.org/doi/10.1103/PhysRevLett.84.2726}.

\bibitem[{\citenamefont{Duan et~al.}(2000)\citenamefont{Duan, Giedke, Cirac,
  and Zoller}}]{PhysRevLett.84.2722}
\bibinfo{author}{\bibfnamefont{L.-M.} \bibnamefont{Duan}},
  \bibinfo{author}{\bibfnamefont{G.}~\bibnamefont{Giedke}},
  \bibinfo{author}{\bibfnamefont{J.~I.} \bibnamefont{Cirac}}, \bibnamefont{and}
  \bibinfo{author}{\bibfnamefont{P.}~\bibnamefont{Zoller}},
  \bibinfo{journal}{Phys. Rev. Lett.} \textbf{\bibinfo{volume}{84}},
  \bibinfo{pages}{2722} (\bibinfo{year}{2000}),
  \urlprefix\url{https://link.aps.org/doi/10.1103/PhysRevLett.84.2722}.

\bibitem[{\citenamefont{Mancini et~al.}(2002)\citenamefont{Mancini,
  Giovannetti, Vitali, and Tombesi}}]{PhysRevLett.88.120401}
\bibinfo{author}{\bibfnamefont{S.}~\bibnamefont{Mancini}},
  \bibinfo{author}{\bibfnamefont{V.}~\bibnamefont{Giovannetti}},
  \bibinfo{author}{\bibfnamefont{D.}~\bibnamefont{Vitali}}, \bibnamefont{and}
  \bibinfo{author}{\bibfnamefont{P.}~\bibnamefont{Tombesi}},
  \bibinfo{journal}{Phys. Rev. Lett.} \textbf{\bibinfo{volume}{88}},
  \bibinfo{pages}{120401} (\bibinfo{year}{2002}),
  \urlprefix\url{https://link.aps.org/doi/10.1103/PhysRevLett.88.120401}.

\bibitem[{\citenamefont{Raymer et~al.}(2003)\citenamefont{Raymer, Funk,
  Sanders, and de~Guise}}]{PhysRevA.67.052104}
\bibinfo{author}{\bibfnamefont{M.~G.} \bibnamefont{Raymer}},
  \bibinfo{author}{\bibfnamefont{A.~C.} \bibnamefont{Funk}},
  \bibinfo{author}{\bibfnamefont{B.~C.} \bibnamefont{Sanders}},
  \bibnamefont{and} \bibinfo{author}{\bibfnamefont{H.}~\bibnamefont{de~Guise}},
  \bibinfo{journal}{Phys. Rev. A} \textbf{\bibinfo{volume}{67}},
  \bibinfo{pages}{052104} (\bibinfo{year}{2003}),
  \urlprefix\url{https://link.aps.org/doi/10.1103/PhysRevA.67.052104}.

\bibitem[{\citenamefont{Hillery and Zubairy}(2006)}]{PhysRevLett.96.050503}
\bibinfo{author}{\bibfnamefont{M.}~\bibnamefont{Hillery}} \bibnamefont{and}
  \bibinfo{author}{\bibfnamefont{M.~S.} \bibnamefont{Zubairy}},
  \bibinfo{journal}{Phys. Rev. Lett.} \textbf{\bibinfo{volume}{96}},
  \bibinfo{pages}{050503} (\bibinfo{year}{2006}),
  \urlprefix\url{https://link.aps.org/doi/10.1103/PhysRevLett.96.050503}.

\bibitem[{\citenamefont{Morin et~al.}(2014)\citenamefont{Morin, Huang, Liu,
  Le~Jeannic, Fabre, and Laurat}}]{morin_remote_2014}
\bibinfo{author}{\bibfnamefont{O.}~\bibnamefont{Morin}},
  \bibinfo{author}{\bibfnamefont{K.}~\bibnamefont{Huang}},
  \bibinfo{author}{\bibfnamefont{J.}~\bibnamefont{Liu}},
  \bibinfo{author}{\bibfnamefont{H.}~\bibnamefont{Le~Jeannic}},
  \bibinfo{author}{\bibfnamefont{C.}~\bibnamefont{Fabre}}, \bibnamefont{and}
  \bibinfo{author}{\bibfnamefont{J.}~\bibnamefont{Laurat}},
  \bibinfo{journal}{Nature Photonics} \textbf{\bibinfo{volume}{8}},
  \bibinfo{pages}{570} (\bibinfo{year}{2014}), ISSN \bibinfo{issn}{1749-4885,
  1749-4893}, \urlprefix\url{http://www.nature.com/articles/nphoton.2014.137}.

\bibitem[{\citenamefont{Gouzien et~al.}(2020)\citenamefont{Gouzien, Brunel,
  Tanzilli, and D'Auria}}]{gouzien2020hybrid}
\bibinfo{author}{\bibfnamefont{{\'E}.}~\bibnamefont{Gouzien}},
  \bibinfo{author}{\bibfnamefont{F.}~\bibnamefont{Brunel}},
  \bibinfo{author}{\bibfnamefont{S.}~\bibnamefont{Tanzilli}}, \bibnamefont{and}
  \bibinfo{author}{\bibfnamefont{V.}~\bibnamefont{D'Auria}},
  \bibinfo{journal}{arXiv preprint arXiv:2002.04450}  (\bibinfo{year}{2020}).

\bibitem[{\citenamefont{Jeong et~al.}(2014)\citenamefont{Jeong, Zavatta, Kang,
  Lee, Costanzo, Grandi, Ralph, and Bellini}}]{jeong_generation_2014}
\bibinfo{author}{\bibfnamefont{H.}~\bibnamefont{Jeong}},
  \bibinfo{author}{\bibfnamefont{A.}~\bibnamefont{Zavatta}},
  \bibinfo{author}{\bibfnamefont{M.}~\bibnamefont{Kang}},
  \bibinfo{author}{\bibfnamefont{S.-W.} \bibnamefont{Lee}},
  \bibinfo{author}{\bibfnamefont{L.~S.} \bibnamefont{Costanzo}},
  \bibinfo{author}{\bibfnamefont{S.}~\bibnamefont{Grandi}},
  \bibinfo{author}{\bibfnamefont{T.~C.} \bibnamefont{Ralph}}, \bibnamefont{and}
  \bibinfo{author}{\bibfnamefont{M.}~\bibnamefont{Bellini}},
  \bibinfo{journal}{Nature Photonics} \textbf{\bibinfo{volume}{8}},
  \bibinfo{pages}{564} (\bibinfo{year}{2014}), ISSN \bibinfo{issn}{1749-4885,
  1749-4893}, \urlprefix\url{http://www.nature.com/articles/nphoton.2014.136}.

\bibitem[{\citenamefont{Fang et~al.}(2014)\citenamefont{Fang, Cohen, and
  Lorenz}}]{FangPola}
\bibinfo{author}{\bibfnamefont{B.}~\bibnamefont{Fang}},
  \bibinfo{author}{\bibfnamefont{O.}~\bibnamefont{Cohen}}, \bibnamefont{and}
  \bibinfo{author}{\bibfnamefont{V.~O.} \bibnamefont{Lorenz}},
  \bibinfo{journal}{J. Opt. Soc. Am. B} \textbf{\bibinfo{volume}{31}},
  \bibinfo{pages}{277} (\bibinfo{year}{2014}),
  \urlprefix\url{http://josab.osa.org/abstract.cfm?URI=josab-31-2-277}.

\bibitem[{\citenamefont{Leonhardt}(1993)}]{Leonhardt1993}
\bibinfo{author}{\bibfnamefont{U.}~\bibnamefont{Leonhardt}},
  \bibinfo{journal}{Phys. Rev. A} \textbf{\bibinfo{volume}{48}},
  \bibinfo{pages}{3265} (\bibinfo{year}{1993}).

\bibitem[{\citenamefont{Kreis}(2012)}]{kreis_characterizing_2012}
\bibinfo{author}{\bibfnamefont{K.}~\bibnamefont{Kreis}},
  \bibinfo{journal}{arXiv:1211.2880 [quant-ph]}  (\bibinfo{year}{2012}),
  \bibinfo{note}{arXiv: 1211.2880},
  \urlprefix\url{http://arxiv.org/abs/1211.2880}.

\bibitem[{\citenamefont{Wolf and Perez-Garcia}(2007)}]{wolf2007quantum}
\bibinfo{author}{\bibfnamefont{M.~M.} \bibnamefont{Wolf}} \bibnamefont{and}
  \bibinfo{author}{\bibfnamefont{D.}~\bibnamefont{Perez-Garcia}},
  \bibinfo{journal}{Physical Review A} \textbf{\bibinfo{volume}{75}},
  \bibinfo{pages}{012303} (\bibinfo{year}{2007}).

\bibitem[{\citenamefont{van Enk and Hirota}(2001)}]{van2001entangled}
\bibinfo{author}{\bibfnamefont{S.~J.} \bibnamefont{van Enk}} \bibnamefont{and}
  \bibinfo{author}{\bibfnamefont{O.}~\bibnamefont{Hirota}},
  \bibinfo{journal}{Physical Review A} \textbf{\bibinfo{volume}{64}},
  \bibinfo{pages}{022313} (\bibinfo{year}{2001}).

\bibitem[{\citenamefont{Wang}(2001)}]{wang2001bipartite}
\bibinfo{author}{\bibfnamefont{X.}~\bibnamefont{Wang}},
  \bibinfo{journal}{Journal of Physics A: Mathematical and General}
  \textbf{\bibinfo{volume}{35}}, \bibinfo{pages}{165} (\bibinfo{year}{2001}).

\bibitem[{\citenamefont{Santos et~al.}(2006)\citenamefont{Santos, Milman,
  Davidovich, and Zagury}}]{santos2006direct}
\bibinfo{author}{\bibfnamefont{M.~F.} \bibnamefont{Santos}},
  \bibinfo{author}{\bibfnamefont{P.}~\bibnamefont{Milman}},
  \bibinfo{author}{\bibfnamefont{L.}~\bibnamefont{Davidovich}},
  \bibnamefont{and} \bibinfo{author}{\bibfnamefont{N.}~\bibnamefont{Zagury}},
  \bibinfo{journal}{Physical Review A} \textbf{\bibinfo{volume}{73}},
  \bibinfo{pages}{040305} (\bibinfo{year}{2006}).

\bibitem[{\citenamefont{Morin et~al.}(2013)\citenamefont{Morin, Bancal, Ho,
  Sekatski, D’Auria, Gisin, Laurat, and Sangouard}}]{morin_witnessing_2013}
\bibinfo{author}{\bibfnamefont{O.}~\bibnamefont{Morin}},
  \bibinfo{author}{\bibfnamefont{J.-D.} \bibnamefont{Bancal}},
  \bibinfo{author}{\bibfnamefont{M.}~\bibnamefont{Ho}},
  \bibinfo{author}{\bibfnamefont{P.}~\bibnamefont{Sekatski}},
  \bibinfo{author}{\bibfnamefont{V.}~\bibnamefont{D’Auria}},
  \bibinfo{author}{\bibfnamefont{N.}~\bibnamefont{Gisin}},
  \bibinfo{author}{\bibfnamefont{J.}~\bibnamefont{Laurat}}, \bibnamefont{and}
  \bibinfo{author}{\bibfnamefont{N.}~\bibnamefont{Sangouard}},
  \bibinfo{journal}{Physical Review Letters} \textbf{\bibinfo{volume}{110}}
  (\bibinfo{year}{2013}), ISSN \bibinfo{issn}{0031-9007, 1079-7114},
  \urlprefix\url{https://link.aps.org/doi/10.1103/PhysRevLett.110.130401}.

\bibitem[{\citenamefont{Qi et~al.}(2020)\citenamefont{Qi, Lougovski, and
  Williams}}]{qi2020characterizing}
\bibinfo{author}{\bibfnamefont{B.}~\bibnamefont{Qi}},
  \bibinfo{author}{\bibfnamefont{P.}~\bibnamefont{Lougovski}},
  \bibnamefont{and} \bibinfo{author}{\bibfnamefont{B.~P.}
  \bibnamefont{Williams}}, \bibinfo{journal}{Optics Express}
  \textbf{\bibinfo{volume}{28}}, \bibinfo{pages}{2276} (\bibinfo{year}{2020}).

\bibitem[{\citenamefont{Le~Jeannic}(2016)}]{lejeannic:tel-01665496}
\bibinfo{author}{\bibfnamefont{H.}~\bibnamefont{Le~Jeannic}},
  \bibinfo{type}{Theses}, \bibinfo{school}{{Universit{\'e} Pierre et Marie
  Curie - Paris VI}} (\bibinfo{year}{2016}),
  \urlprefix\url{https://tel.archives-ouvertes.fr/tel-01665496}.

\bibitem[{\citenamefont{Laiho et~al.}(2009)\citenamefont{Laiho, Avenhaus,
  Cassemiro, and Silberhorn}}]{laiho2009direct}
\bibinfo{author}{\bibfnamefont{K.}~\bibnamefont{Laiho}},
  \bibinfo{author}{\bibfnamefont{M.}~\bibnamefont{Avenhaus}},
  \bibinfo{author}{\bibfnamefont{K.}~\bibnamefont{Cassemiro}},
  \bibnamefont{and}
  \bibinfo{author}{\bibfnamefont{C.}~\bibnamefont{Silberhorn}},
  \bibinfo{journal}{New Journal of Physics} \textbf{\bibinfo{volume}{11}},
  \bibinfo{pages}{043012} (\bibinfo{year}{2009}).

\bibitem[{\citenamefont{Sridhar et~al.}(2014)\citenamefont{Sridhar,
  Shahrokhshahi, Miller, Calkins, Gerrits, Lita, Nam, and
  Pfister}}]{sridhar2014direct}
\bibinfo{author}{\bibfnamefont{N.}~\bibnamefont{Sridhar}},
  \bibinfo{author}{\bibfnamefont{R.}~\bibnamefont{Shahrokhshahi}},
  \bibinfo{author}{\bibfnamefont{A.~J.} \bibnamefont{Miller}},
  \bibinfo{author}{\bibfnamefont{B.}~\bibnamefont{Calkins}},
  \bibinfo{author}{\bibfnamefont{T.}~\bibnamefont{Gerrits}},
  \bibinfo{author}{\bibfnamefont{A.}~\bibnamefont{Lita}},
  \bibinfo{author}{\bibfnamefont{S.~W.} \bibnamefont{Nam}}, \bibnamefont{and}
  \bibinfo{author}{\bibfnamefont{O.}~\bibnamefont{Pfister}},
  \bibinfo{journal}{JOSA B} \textbf{\bibinfo{volume}{31}}, \bibinfo{pages}{B34}
  (\bibinfo{year}{2014}).

\end{thebibliography}

\clearpage
\appendix
\onecolumngrid

\section{Thermal Noise}
\label{Thermalnoise}
We study the effect of adding thermal photon noise. To do so, we replace the vacuum that we put on the beam-splitter of the continuous channel by a thermal state, as featured in Figure \ref{schemeofsetupTS}.

\begin{centering}
\begin{figure}
	\begin{tikzpicture}[thick, scale=0.8]
\draw(10,1)--(10,2) ; 
	\draw(0,2)--(12,2) ; 
	\draw(9,1)--(11,3) ; 
	\draw(7,3)--(5,1) ; 
	\draw(6,2)--(6,8) ; 
	\draw(6,5)--(7,5) ; 
	\draw(0,8)--(12,8) ; 
	\draw(5,9)--(7,7) ; 
	\draw(9,9)--(11,7) ; 
	\draw(10,9)--(10,8) ; 
	\draw (10,1) node[below]{thermal state} ;
	\draw (0,2) node[below]{Photon Pair} ;
	\draw (12,2) node[right]{Homodyne detector} ;
	\draw (12,8) node[right]{Homodyne detector} ;
	\draw (6,2) node[above left]{PBS} ;
	\draw (10,2) node[above left]{TBS} ;
	\draw (10,8) node[below left]{TBS} ;
	\draw (7,5) node[right]{Photon detector} ;
	\draw (10,9) node[above]{thermal state} ;
	\draw (6,8) node[below left]{BS} ;
	\draw (0,8) node[above]{Squeezed vacuum} ;
	\end{tikzpicture}
	\caption{A scheme of the set-up, with the theoretical beam splitters (TBS) which purpose are to take into account noise in the set up}
	\label{schemeofsetupTS}
\end{figure}
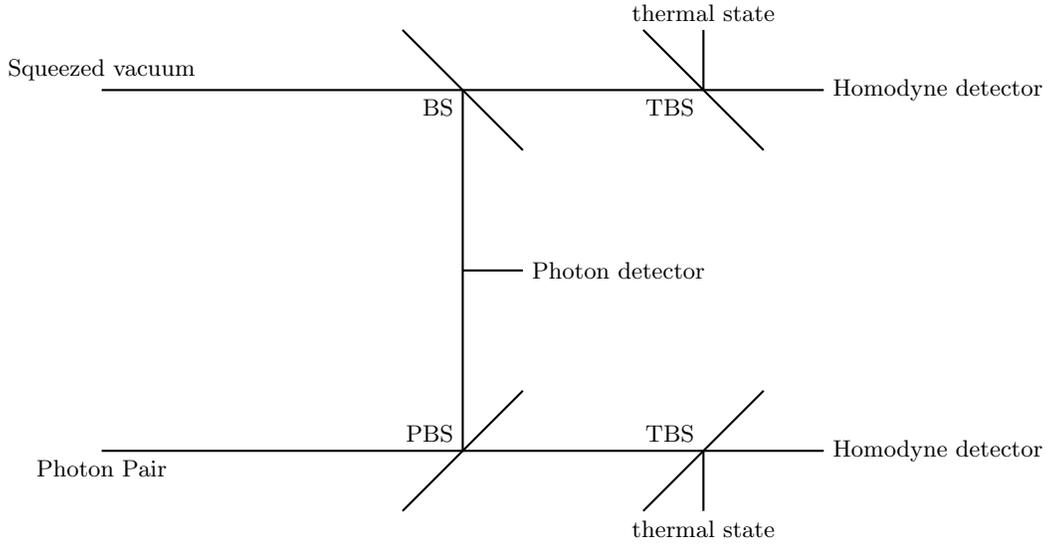
\end{centering}

We derive again the value of the Witness. The effect of the thermal noise on coherent states can be written, in agreement with Kreis and Van Loock \cite{kreis_classifying_2012}: 
\begin{align*}
   \$_{\text{thermal}} \left( \ket{\alpha} \bra{\alpha} \right)
  &= \frac{1}{2 \pi \moy{\text{nth}}} \int_{\mathbb{C}} \mathrm{d}^2 \gamma \exp{ \left( \frac{- |\gamma|^2}{\moy{\text{nth}}} \right)} \ket{\sqrt{1-\eta} \alpha - \sqrt{\eta} \gamma} \bra{\sqrt{1-\eta} \alpha - \sqrt{\eta} \gamma} \\
      &=   \frac{1}{2 \pi \moy{\text{nth}}} \int_{\mathbb{C}} \mathrm{d}^2 \gamma   \exp{ \left( \frac{- |\gamma|^2}{\moy{\text{nth}}} \right)} \mathbb{D} \left( \sqrt{1-\eta} \alpha  \right) \ket{\sqrt{\eta} \gamma }  \bra{\sqrt{\eta} \gamma }  \mathbb{D}^{\dagger} \left( \sqrt{1-\eta} \alpha \right) \\
      &=   \frac{1}{2 \pi \moy{\text{nth}}} \mathbb{D} \left( \sqrt{1-\eta} \alpha  \right)  \left( \int_{\mathbb{C}} \mathrm{d}^2 \gamma   \exp{ \left( \frac{- |\gamma|^2}{\moy{\text{nth}}} \right)} \ket{\sqrt{\eta} \gamma }  \bra{\sqrt{\eta} \gamma }  \right) \mathbb{D}^{\dagger} \left( \sqrt{1-\eta} \alpha \right) \\
      &= \mathbb{D} \left( \sqrt{1-\eta} \alpha \right)    \$_{\text{thermal}} \left( \ket{0} \bra{0} \right)   \mathbb{D}^{\dagger} \left(\sqrt{1-\eta} \alpha \right) 
\end{align*}
where $\mathbb{D}$ is the displacement operator such that $\mathbb{D} (\alpha) \ket{0} = \ket{\alpha}$.
From this, we can compute again the value of the witness by computing the effect of the thermal noise on the cat states.
\par An interesting point is that the mean value of linear combinations of ladder operators with our thermal state does not depend on the temperature. \textit{This implies that the mean value of the witness defined in equation~\eqref{finalwitness} for our class of noisy state is the same at all temperatures}.
\par However, we still have to take into account thermal effects in the boundary of the value of the witness for separable state given in Appendix \ref{appendixwitness}. The number of photons in the continuous part can only grow, and the critical value of the maximum number of photons for which our witness is still positive for separable state would grow. The quantification of the number of additional photons is obtained through the formula :  $\moy{n} = \frac{1}{e^{\frac{h \nu}{k_B T}} - 1}$.

\clearpage
\section{Concurrence}
\label{ConcurrenceAppendix}

The expression of $\rho_{\text{full noise}}$ is in the basis formed by the kets 
\begin{equation}
    \ket{0}, \ket{1}, \ket{\sqrt{1-\eta}\alpha}, \ket{- \sqrt{1-\eta}\alpha}
\end{equation}

\begin{align*}
    \rho_{\text{noise}} =  &\ket{0}\bra{0} 
    \left[ \ket{\sqrt{1-\eta}\alpha} \bra{\sqrt{1-\eta}\alpha} \left( \frac{1}{2 N^{-2}} + \frac{\eta_d}{2 N^{+2}} \right)  - \ket{\sqrt{1-\eta}\alpha} \bra{- \sqrt{1-\eta}\alpha} \left( \frac{(\eta_d) f(\eta)}{2 N^{+2}} + \frac{\eta_d}{2 N^{-2}} \right) \right. \\
    &\phantom{\ket{0}\bra{0}}
     + \left. \ket{-\sqrt{1-\eta}\alpha} \bra{\sqrt{1-\eta}\alpha} \left( \frac{- f(\eta)}{2 N^{-2}} + \frac{\eta_d f(\eta)}{2 N^{-2}} \right)  + \ket{-\sqrt{1-\eta}\alpha} \bra{- \sqrt{1-\eta}\alpha} \left( \frac{\eta_d}{2 N^{+2}} + \frac{\eta_d}{2 N^{+2}} \right)  \right] \\
    + &\ket{1}\bra{1} \left[ \ket{\sqrt{1-\eta}\alpha} \bra{\sqrt{1-\eta}\alpha}   \frac{1-\eta_d}{2 N^{+2}}   + \ket{\sqrt{1-\eta}\alpha} \bra{- \sqrt{1-\eta}\alpha}  \frac{(1-\eta_d) f(\eta)}{2 N^{+2}}  \right. \\
    &\phantom{\ket{0}\bra{0}}
     + \left. \ket{-\sqrt{1-\eta}\alpha} \bra{\sqrt{1-\eta}\alpha}  \frac{1-\eta_d f(\eta)}{2 N^{+2}}   + \ket{-\sqrt{1-\eta}\alpha} \bra{- \sqrt{1-\eta}\alpha} \frac{1-\eta_d}{2 N^{+2}}
    \right] \\
    + &\ket{0}\bra{1} \left[ \ket{\sqrt{1-\eta}\alpha} \bra{\sqrt{1-\eta}\alpha}   \frac{\sqrt{1-\eta_d}}{2 N^+N^-}   + \ket{\sqrt{1-\eta}\alpha} \bra{- \sqrt{1-\eta}\alpha}  \frac{\sqrt{1-\eta_d} f(\eta)}{2 N^-N^+}  \right. \\
    &\phantom{\ket{0}\bra{0}}
      \left. 
     - \ket{-\sqrt{1-\eta}\alpha} \bra{\sqrt{1-\eta}\alpha}  \frac{\sqrt{1-\eta_d} f(\eta)}{2 N^+N^-} - \ket{-\sqrt{1-\eta}\alpha} \bra{- \sqrt{1-\eta}\alpha} \frac{\sqrt{1-\eta_d}}{2 N^+N^-}
    \right] \\
    + &\ket{1}\bra{0} \left[ \ket{\sqrt{1-\eta}\alpha} \bra{\sqrt{1-\eta}\alpha}   \frac{\sqrt{1-\eta_d}}{2 N^+N^-}   - \ket{\sqrt{1-\eta}\alpha} \bra{- \sqrt{1-\eta}\alpha}  \frac{\sqrt{1-\eta_d} f(\eta)}{2 N^-N^+}  \right. \\
    &\phantom{\ket{0}\bra{0}}
     + \left. 
     \ket{-\sqrt{1-\eta}\alpha} \bra{\sqrt{1-\eta}\alpha}  \frac{\sqrt{1-\eta_d} f(\eta)}{2 N^+N^-} - \ket{-\sqrt{1-\eta}\alpha} \bra{- \sqrt{1-\eta}\alpha} \frac{\sqrt{1-\eta_d}}{2 N^+N^-}
    \right] 
\end{align*}
 with $ f(\eta) = \exp{(- 2 \eta \alpha^2)}$ and $N^{\pm} = N^{\pm}(\alpha)$ (see Eq~\eqref{eq:catStateDef} of the main text).


If we take a specific orthonormalisation, that of equations \ref{dampedcat}, the matrix is written 
\begin{equation}
   \rho_{\text{full noise}} = \begin{pmatrix}
         w & 0 & 0 & z  \\
         0 & x_1 & c & 0 \\
        0 & c & x_2 & 0 \\
        z & 0 & 0 & y 
        \label{densitynoisymatrix}
       \end{pmatrix} ,
\end{equation}
with 
\begin{align}
    &w \left (\eta, \eta_d, \alpha \right) = \frac{(1 + f(1-\eta, 
      \alpha))\eta_d (1 + f(\eta, \alpha))}{
       N^+(\alpha)^2} + \frac{(1 - f(\eta, \alpha))}{2 N^-(\alpha)^2} \\
    &z(\eta, \eta_d, \alpha) = \frac{\sqrt{1 + f(1-\eta, \alpha)} \sqrt{1 - 
    f(1-\eta, \alpha))}\sqrt{1-\eta} (1 - f(\eta, \alpha))}{N^-(\alpha) N^+(\alpha)} \\
    &x_1(\eta, \eta_d, \alpha) = \frac{(1 + f(1-\eta, \alpha)) (1-\eta_d) \left(1 + f(\eta, \alpha) \right) }{2 N^+(\alpha)^2} \\
    &x_2(\eta, \eta_d, \alpha) = \frac{(1 - f(1-\eta, \alpha)) (1 + f(\eta, \alpha))}{2 N^-(\alpha)^2} + \frac{\eta_d (1 - f(\eta, \alpha))}{2  N^+(\alpha)^2} \\
   &c(\eta, \eta_d, \alpha) = \frac{\sqrt{1 + f(1-\eta, \alpha)}\sqrt{1 - 
   f(1-\eta, \alpha)}  \sqrt{1-\eta_d} (1 + f(\eta, 
   \alpha))}{2 N^-(\alpha) N^+(\alpha)} \\
  & y(\eta, \eta_d, \alpha)= \frac{\left(1 - f(1-\eta, \alpha) \right)  (1-\eta_d)  \left(1 - f(\eta, \alpha)\right)}{2 N^+(\alpha)^2}
\end{align}

\clearpage
\section{Proof that \texorpdfstring{$\tilde{W}$}{W} is a witness}
\label{appendixwitness}

Let us prove that $\tilde{W}$, as defined in equation \eqref{finalwitness} is an Entanglement Witness. 

\begin{equation}
     \tilde{W} = \mathds{1} - \frac{1}{2}\left[ \mathds{1} + X_D \otimes \frac{a+a^{\dagger}}{n_x} - Y_D \otimes  \frac{i(a-a^{\dagger})}{n_y}  \right]
\end{equation}

We need to check that 

\begin{equation}
     \tilde{W}_{\sigma} \geq 0 \ \forall \ \sigma \ \text{separable}
\end{equation}
This is equivalent to proving that 
\begin{equation}
    \moy{X_D \otimes \frac{a+a^{\dagger}}{n_x} - Y_D \otimes  \frac{i(a-a^{\dagger})}{n_y}}_{\sigma} \leq 1 \ \forall \ \sigma \ \text{separable}
    \label{tobeprovenEW}
\end{equation}
The most general separable state can be written: 
\begin{equation}
   \sigma = \sum_k \lambda_k \sigma_D \otimes \sigma_C
\end{equation}
with  $\{ \lambda_k \} $ being a convex set, $\sigma_D $ is a $2 \times 2$ matrix and $\sigma_C$ a $N \times N$ matrix that feature the continuous part of the state, $N$ being the cut-off of the Hilbert space of the continuous part.

We present the proof of equation (\ref{tobeprovenEW}) with a pure state on the continuous part, the generalisation to mixed states being obtained by convexity. Let 
\begin{equation}
   \tilde{\sigma}  = \sigma_D \otimes \ket{\psi}\bra{\psi}
\end{equation}
with 
\begin{equation}
    \ket{\psi} = \sum_{i=0}^N \lambda_i \ket{i}
\end{equation}
with $\ket{i}$ an eigenstate of operators $a$ and $a^{\dagger}$, with $ \sum_{i=0}^N |\lambda_i|^2$
Since $\sigma_D$ is a density matrix, we write it
\begin{equation}
\sigma_D = \pmat{\sigma_{11}}{\sigma_{12}}{\bar{\sigma}_{12}}{1-\sigma_{11}} 
\end{equation}
with $0 \leq \sigma_{11} \leq 1$ and the following inequality hold : 
\begin{equation}
    \det (\sigma_D ) \geq 0 \iff |\sigma_{12}| \leq \sqrt{\sigma_{11}(1-\sigma_{11})}
    \label{determdensity}
\end{equation}
Now, 
\begin{equation}
    \moy{X_D \otimes \frac{a+a^{\dagger}}{n_x}}_{\Tilde{\sigma}} = (\sigma_{12}+\bar{\sigma}_{12} \left( \sum_{i=0}^{N-1} (\bar{\lambda}_i \lambda_{i+1} \sqrt{i+1}+ \lambda_i \bar{\lambda}_{i+1} \sqrt{i+1} )\times \frac{1}{n_x} \right)
\end{equation}

\begin{equation}
    \moy{Y_D \otimes \frac{i (a-a^{\dagger})}{n_y}}_{\Tilde{\sigma}} =   (\sigma_{12}-\bar{\sigma}_{12})\left( \sum_{i=0}^{N-1} (\bar{\lambda}_i \lambda_{i+1} \sqrt{i+1} - \lambda_i \bar{\lambda}_{i+1} \sqrt{i+1} )\times \frac{1}{n_y} \right)
\end{equation}
Hence, 
\begin{align}
  \left|\moy{X_D \otimes \frac{a+a^{\dagger}}{n_x}}_{\Tilde{\sigma}} - \moy{Y_D \otimes \frac{i (a-a^{\dagger})}{n_y}}_{\Tilde{\sigma}}\right| 
  &= \left| (\sigma_{12}+\bar{\sigma_{12}})\left( \sum_{i=0}^{N-1} (\bar{\lambda}_i \lambda_{i+1} \sqrt{i+1}+ \lambda_i \bar{\lambda}_{i+1} \sqrt{i+1} )\times \frac{1}{n_x} \right) \right.\\ 
  &\hphantom{=|}- \left. i (\sigma_{12}-\bar{\sigma_{12}})\left( \sum_{i=0}^{N-1} (\bar{\lambda}_i \lambda_{i+1} \sqrt{i+1} - \lambda_i \bar{\lambda}_{i+1} \sqrt{i+1} )\times \frac{1}{n_y} \right) \right| \\
  &= \left| \frac{2 \Re(\sigma_{12}) 2 \Re(\sum_{i=0}^{N-1} (\bar{\lambda}_i \lambda_{i+1} \sqrt{i+1})}{n_x} - \frac{2 \Im(\sigma_{12}) 2 \Im(\sum_{i=0}^{N-1} (\bar{\lambda}_i \lambda_{i+1} \sqrt{i+1})}{n_y} \right| \\
      &= 4 |\sigma_{12}| \left| \frac{ \cos(\sigma_{12})  \sum_{i=0}^{N-1} \Re(\bar{\lambda}_i \lambda_{i+1} \sqrt{i+1})}{n_x}  - \frac{\sin(\sigma_{12})  \sum_{i=0}^{N-1} \Im(\bar{\lambda}_i \lambda_{i+1} \sqrt{i+1})}{n_y} \right| \\
  &= 4 |\sigma_{12}| \left| \frac{ \cos(\sigma_{12})  \sum_{i=0}^{N-1} |\bar{\lambda}_i| |\lambda_{i+1}|\sqrt{i+1}) \cos(\theta_{k})}{n_x}\right.\\
  &\hphantom{= 4 |\sigma_{12}||}\left.- \frac{\sin(\sigma_{12})  \sum_{i=0}^{N-1} |\bar{\lambda}_i|| \lambda_{i+1} | \sqrt{i+1}) \sin(\theta_{k}) }{n_y} \right| \\
&= 4 |\sigma_{12}| \left| \sum_{i=0}^{N-1} |\bar{\lambda}_i| |\lambda_{i+1}|\sqrt{i+1}) \left( \frac{ \cos(\sigma_{12})   \cos(\theta_{k} )}{n_x} - \frac{\sin(\sigma_{12}) \sin(\theta_{k}) }{n_y} \right) \right| \\
\end{align}
where we have used $\lambda_k = | \lambda_k | \exp{i \phi_k}$ and $\theta_{k} = \phi_{k+1} - \phi_k$
Finally, 
\begin{align}
        |\moy{X_D \otimes \frac{a+a^{\dagger}}{n_x}}_{\Tilde{\sigma}} - \moy{Y_D \otimes \frac{i (a-a^{\dagger})}{n_y}}_{\Tilde{\sigma}}| &\leq 2 \left| \sum_{i=0}^{N-1} |\bar{\lambda_i}| |\lambda_{i+1}|\sqrt{i+1}) \max(\frac{1}{|n_x|},\frac{1}{|n_y|})  \right| \\
        & \leq 2 \left| f(N) \max(\frac{1}{|n_x|},\frac{1}{|n_y|})  \right|
\end{align}
where we have used $\underset{\sigma_{11} \in [0;1]}{\max} \sqrt{\sigma_{11}(1-\sigma_{11})} = \frac{1}{2}$, $\left| \frac{ \cos(\sigma_{12})   \cos(\theta_{k} )}{n_x}  - \frac{\sin((\sigma_{12}) \sin(\theta_{k}) }{n_y} \right| \leq \max(\frac{1}{|n_x|},\frac{1}{|n_y|})$
 and defined $ f(N) = \sup_{\{ \lambda_i \}} \sum_{i=0}^{N-1} |\lambda_i | |\bar{\lambda}_{i+1} ) | \sqrt{i+1} )$ with the constraint that the set of $\lambda_i$ is convex.  

This upper bound depends on the number of photons since it is related to the truncation in the Fock basis. Now, we can either take an arbitrary cut-off and compute a worst-case scenario with an optimisation problem, or we can upper bound it tightly if we possess the knowledge of the set of \{$\lambda_i$\}. The former is presented below whereas the latter procedure, to obtain \{$\lambda_i$\} and upper bound the terms is explained in Appendix \ref{Control}.

\subsection*{Method 1: worst-case scenario}

We write extensively the optimisation problem in equation \eqref{systemoptimize}.

\begin{equation}
  \left\{
      \begin{aligned}
       \sum_{i=0}^{N-1} |\lambda_i| |\bar{\lambda}_{i+1}|   \sqrt{i+1} \\
       \sum_{i=0}^{N} \left| \lambda_i \right|^2 =1
      \end{aligned}
    \right.
    \label{systemoptimize}
\end{equation}
Let's write $\lambda_k = |\lambda_k| e^{i \phi_k}$. Now, $\Re(\lambda_k \bar{\lambda}_{k+1}) = |\lambda_k| |\lambda_{k+1}|\cos (\phi_k - \phi_{k+1} )$. If we want to maximise this quantity, we need $\phi_k - \phi_{k+1} = 2 n \pi$. This constrain can indeed be reached. Without lost of generality, we can consequently consider that the $\lambda_k$ are positive reals. This hypothesis is done from now on. 

If we use the formalism of Lagrange multipliers, we need to nullify the gradient of 

 \begin{equation}
 \mathcal{L} (\lambda_k,\lambda) = f(\lambda_k) - \mu g(\lambda_k)
\end{equation}
with respect to every $\lambda_k$ and $\mu$, with 
\begin{equation}
  \left\{
      \begin{aligned}
       f(\lambda_k) = \sum_{i=0}^{N-1} |\lambda_i| |\lambda_{i+1}| \sqrt{i+1} \\
       g(\lambda_k) = 1 - \sum_{i=0}^{N} \left| \lambda_i \right|^2 
      \end{aligned}
    \right.
\end{equation}

We obtain : 
\begin{align}
  \left\{
      \begin{aligned}
       \frac{\partial  \mathcal{L} }{ \partial \mu} = - \left( 1 - \sum_{i=0}^{N} \left| \lambda_i \right|^2  \right) = 0  \\
        \forall \ j \in [1;N-1] \ \ \frac{\partial  \mathcal{L} }{ \partial \lambda_k}  = x_{j+1} \sqrt{j+1}+x_{j-1} \sqrt{j} + 2 \mu x_j   = 0    \\
        \frac{\partial  \mathcal{L} }{ \partial \lambda_0}  = x_{1} \sqrt{1} + 2 \mu x_0 = 0  \\
        \frac{\partial  \mathcal{L} }{ \partial \lambda_N} = x_{N-1} \sqrt{N} + 2 \mu x_N = 0 
      \end{aligned}
    \right.
    \label{Lagrangedivise}
\end{align}
We can recast the last three lines into the following linear problem : 
\begin{equation}
    \begin{pmatrix}
   2 \mu & \sqrt{1} & \cdots \\ 
   \sqrt{1} & 2 \mu & \sqrt{2} & \cdots \\
    \vdots & \ddots & \vdots \\ 
    \cdots & \sqrt{N-1} & 2 \mu & \sqrt{N} \\
    \cdots & ... & \sqrt{N} & 2 \mu 
\end{pmatrix}
\begin{pmatrix} x_0 \\ x_1 \\ \vdots \\ x_{N-1} \\ x_N \end{pmatrix} = 0
\label{LinearProblem}
\end{equation}
Now, we have 
\begin{equation}
  M(N) \vec{x} = \vec{0}.
\end{equation}
where $M$ is the matrix displayed in \eqref{LinearProblem} and $\vec{x}$ the vector.
There exists a solution for 
\begin{equation}
    \det \left(  M(N) \right) = 0
\end{equation}
The determinant of M, which we note $d_N$, follows the recurrence relation : 
\begin{equation}
    d_{N+2} (\mu)= 2 \mu d_{N+1}(\mu) - (N+1) d_N(\mu).
\end{equation}
We can recognise the recurrence relation of the Hermite Polynomials. The protocol to solve the optimization problem is the following : 
\begin{itemize}
    \item For a given $N$, compute the roots of the $Nth$ Hermite Polynomials
    \item For every root, compute the M matrix, and look for the kernel of M
    \item Keep $\vec{x}$ for which all terms have the same sign
    \item Normalise it using the previous Lagrangian constraint
\end{itemize}
\par Since $f(N)$, even after optimisation, is above 1, we will have to damp our witness to ensure it is always positive for separable state. This, in turn, will decrease the number of entangled states that can be detected. We are able to produce a witness for experimentally interesting values; we can obtain negative values for cat states of size 1 and for which the noise $\eta, \eta_d$ can go up to 0.25, provided $N$ is not above 4.

\clearpage
\section{Naimark Extension}
\label{NaimarkAppendix}

We give explicit calculations of the Krauss Operators. We looked for expressions of $U_c$ and $U_d$ through
\begin{equation}
    \rho_{\text{noise}} = Tr_{r_c,r_d} \left( U_c \otimes U_d \right) \left( \rho \otimes \mathcal{R} \right) \left( U_c^{\dagger} \otimes U_d^{\dagger}, \right)
    \label{Naimark}
\end{equation}
with $\mathcal{R}$ a reservoir at zero temperature, 
\begin{equation}
   \mathcal{R} = \begin{pmatrix}
        1 & 0 & 0 & 0  \\
         0 & 0 & 0 & 0 \\
        0 & 0 & 0 & 0 \\
        0 & 0 & 0 & 0 
        \label{densitynoisymatrix2}
       \end{pmatrix} ,
\end{equation} 
and
\begin{equation}
    U_{c,d} = \begin{pmatrix}
         e^{i \beta_1} \cos \left( \delta \right) & 0 & 0 & e^{i \beta_2} \sin \left( \delta \right)  \\
      0 & e^{i \phi_1} \cos \left( \gamma \right) & e^{i \phi_2} \sin \left( \gamma \right) & 0 \\
        0 & -e^{-i \phi_2} \sin \left( \gamma \right) & e^{-i \phi_1} \cos \left( \gamma \right) & 0 \\
        -e^{-i \beta_2} \sin \left( \delta \right) & 0 & 0 & e^{-i \beta_1} \cos \left( \delta \right)
        \label{densitynoisymatrix3}
       \end{pmatrix},
\end{equation}
so that they could modelise exchange of photons between the system and the reservoirs. We recall that $\rho$ does not live in the same exact basis than $\rho_{\text{full noise}}$ which is why our Naimark's transformation is a formal one. 

\begin{equation}
   \rho = \begin{pmatrix}
         0 & 0 & 0 & 0  \\
         0 & \frac{1}{2} & \frac{1}{2} & 0 \\
        0 & \frac{1}{2} & \frac{1}{2} & 0 \\
        0 & 0 & 0 & 0 
        \label{densitynoisymatrix4}
       \end{pmatrix} 
\end{equation}.

In the basis
\begin{equation}
    \ket{+}_{d} \ket{0}_{r_d}, \ket{+}_{d} \ket{1}_{r_d}, \ket{-}_{d} \ket{0}_{r_d}, \ket{-}_{d} \ket{1}_{r_d}, 
\end{equation}
where d stands for discrete and $r_d$ discrete reservoir, the expression of $U_d$ is given by the following matrix.
\begin{equation}
   U_d = \begin{pmatrix}
         1 & 0 & 0 & 0  \\
         0 & \sqrt{1-\eta_d} & \sqrt{\eta_d} & 0 \\
        0 & - \sqrt{\eta_d} & \sqrt{\eta_d} & 0 \\
        0 & 0 & 0 & 1 
        \label{UD}
       \end{pmatrix}.
\end{equation}

On the continuous side, in the basis:
\begin{equation}
    \ket{+}_{c} \ket{0}_{r_c}, \ket{+}_{c} \ket{1}_{r_c}, \ket{-}_{c} \ket{0}_{r_c}, \ket{-}_{c} \ket{1}_{r_c}, 
\end{equation}
where c stands for discrete and $r_c$ continuous reservoir, the expression of $U_c$ is given by:
\begin{equation}
   U_c = \begin{pmatrix}
         \frac{\sqrt{(1+K)(1+f)}}{N^+} & 0 & 0 & \frac{\sqrt{(1-K)(1-f)}}{N^+}  \\
         0 & \frac{\sqrt{(1-K)(1+f)}}{N^-} & \frac{\sqrt{(1+K)(1-f)}}{N^-} & 0 \\
        0 & - \frac{\sqrt{(1+K)(1-f)}}{N^-} & \frac{\sqrt{(1-K)(1+f)}}{N^-} & 0 \\
      -\frac{\sqrt{(1-K)(1-f)}}{N^+} & 0 & 0 & \frac{\sqrt{(1+K)(1+f)}}{N^+} 
        \label{UD1}
       \end{pmatrix} .
\end{equation}
 with $K = \exp{(- 2 (1-\eta) \alpha^2)}, f = \exp{(- 2 \eta \alpha^2)}$.
We obtain 
\begin{equation}
    \rho_{\text{continuous noise}} = Trp_{r_c,r_d} \left( Uc \otimes \mathbf{1} \right) \left( \rho \otimes \mathcal{R} \right) \left( Uc^{\dagger} \otimes\mathbf{1} \right)
    \label{Naimarkc}
\end{equation}
,
\begin{equation}
    \rho_{\text{discrete noise}} = Trp_{r_c,r_d} \left( \mathbf{1} \otimes Ud \right) \left( \rho \otimes \mathcal{R} \right) \left( \mathbf{1} \otimes Ud^{\dagger} \right)
    \label{Naimarkd}
\end{equation}
\begin{equation}
   \rho_{\text{discrete noise}} = \begin{pmatrix}
         \frac{\eta_d}{2} & 0 & 0 & 0  \\
         0 & \frac{1-\eta_d}{2} & \frac{\sqrt{1-\eta_d}}{2} & 0 \\
        0 & \frac{\sqrt{1-\eta_d}}{2} & \frac{1}{2} & 0 \\
        0 & 0 & 0 & 1 
        \label{UD22}
       \end{pmatrix} 
\end{equation}
and $\rho_{\text{continuous noise}}$ equal to $\rho_{\text{noise}}$ defined in \eqref{densitynoisymatrix1}, with $\eta_d$ set to 0 (the shape of the matrix doesn't change). We note that  $\rho_{\text{discrete noise}}$ can be expressed as the convex sum of two pure states 
 $\rho_{\text{ent}}=\ket{\psi_{\text{ent}}}\bra{\psi_{\text{ent}}}$ and $\rho_{\text{sep}} = \ket{\psi_{\text{sep}}}\bra{\psi_{\text{sep}}}$,
\begin{equation}
    \rho_{\text{discrete noise}} = \frac{1+(1-\eta_d)}{2} \rho_{\text{ent}} + \frac{1-(1-\eta_d)}{2} \rho_{\text{sep}}
\end{equation}
with $\ket{\psi_{\text{ent}}} = \frac{\ket{0} \ket{-} + \sqrt{(1-\eta_d)} \ket{1} \ket{+}}{\sqrt{\eta_d}}$ an entangled state, and $\ket{\psi_{\text{sep}}} = \ket{0} \ket{+}$ a separable one. The computations of the Krauss operators, from $U_c$ and $U_d$, is straightforward.

\clearpage
\section{Control \textit{a posteriori}}
\label{Control}

We present here the Method 2 described briefly in the main text. It consists in obtaining the photon number distribution, and then to use this information to obtain a precise bound for the witness described in Appendix \ref{appendixwitness}.

\subsection{Photon number statistics evaluation}
In this section, we make a brief summary of the technique presented in the article \textit{Characterizing photon number statistics using conjugate optical homodyne detection} \cite{qi2020characterizing} to evaluate very precisely the photon number distribution of a given experimental state thanks to two conjugate homodyne detectors. 
Given these apparatus, we measure simultaneously two orthogonal quadratures of the electro magnetic-field on two different modes, which have been separated by a beam-splitter. We shall denote them $\widehat{X}$ and $\widehat{P}$, where
\begin{equation}
    \widehat{X} = \frac{1}{\sqrt{2}} \left( \widehat{a}^{\dagger} \exp{(i \theta)} + \widehat{a} \exp{(- i \theta)} \right)
\end{equation}
\begin{equation}
    \widehat{P} = \frac{i}{\sqrt{2}} \left( \widehat{b}^{\dagger} \exp{(i \theta)} - \widehat{b} \exp{(- i \theta)}. \right)
\end{equation}
$\theta$ is the phase of the local oscillator of one of the homodyne, $\widehat{a}^{\dagger}$ and $\widehat{a}$ are photon creation and annihilation on one mode, $\widehat{b}^{\dagger}$ and $\widehat{b}$ on the other one. 
This allows us to form the following observable
\begin{equation}
    \widehat{Z} = \widehat{X}^2 +\widehat{P}^2.
\end{equation}
We can see it as an approximation of the photon number operator $\widehat{N}$. The probability distribution function $ P_Z(z)$ of this observable depends only on the diagonal terms $\rho_{nn}$ of the density matrix, according to the equation:
\begin{equation}
    P_Z(z) = \exp(-z) \sum_{n=0}^{\infty} \frac{\rho_{nn}}{n!}z^n.
\end{equation}
It means that the photon number distribution can be evaluated without having to scan the phase of the LO. Given a repeated sequence of measurement of $\widehat{Z}$, we obtain $P_Z(z)$. Then, thanks to a Bayes inversion and an algorithm of Maximum Likelihood, we are able to infer the diagonal terms of the density matrix $\rho_{nn}$.

\subsection{Precise upper bounding of the witness for separable states}

 This in turns allow us to upper bound equation (C18) and certify that our witness can not produce false-positives. The computation of Appendix $\ref{appendixwitness}$ is established for pure states but the upper bounding we propose stays true in the general case of mixed states. The calculations carried out in equation (C18) involve the terms of the sup and sub diagonal of the density matrix, but they can be bounded by the diagonal terms, according to the following expression:
\begin{equation}
|\rho_{i,i\pm 1}| \leq \sqrt{|\rho_{i,i}||\rho_{i+1,i+1}|}.
\end{equation}

This relation is an equality in the case of pure states. If we consider a mixed state:
\begin{equation}
    \rho = \sum_i p_i \ket{\psi_i} \bra{\psi_i}
\end{equation}
with 
\begin{equation}
    \ket{\psi_i} = \sum_k \lambda_{k,i} \ket{k},
\end{equation}
we have
\begin{align}
    |\rho_{i,i+1}| &= |\sum_k p_k \lambda_{i,k} \bar{\lambda}_{i+1,k}| \\
                    & \leq \sum_k |p_k \lambda_{i,k} \bar{\lambda}_{i+1,k}| \\
                    & = \sum_k \sqrt{p_k^2 |\lambda_{i,k}|^2 |\bar{\lambda}_{i+1,k}|^2} \\
                    &\leq  \sqrt{\sum_k p_k |\lambda_{i,k}|^2 \sum_l p_l |\lambda_{i+1,l}|^2} \\
                    &= \sqrt{|\rho_{i,i}||\rho_{i+1,i+1}|}.
\end{align}
The additional experimental information we measured makes it possible for us to define a tight upper bound for the separable states, that does not necessitate any \textit{prior} knowledge of the states that are produced. 

\subsection{A protocol with only one homodyne detector}

We present a simple protocol to make a rough evaluation of the photon number distribution $\rho_{nn}$ that only necessitates one homodyne detector, and where the quadratures will not be measured jointly. We proceed in two times, recording successively the values of $\widehat{X}$, and then the values of $\widehat{P}$ for two different set of states. We give indices 1 and 2 to the values obtained for respectively the first set of experiments and the second. Since we can not match anymore the right value of $\widehat{X}$ to that of $\widehat{P}$, we will form 
\begin{equation}
 \moy{\widehat{Z}} = \moy{\widehat{X}^2 +\widehat{P}^2 }  \leq  \moy{\widehat{X}^2_1 +\sup_{\widehat{P}_2}{\widehat{P}^2_2}} = \moy{\widehat{Z}_{sup}}.
 \label{Zsup}
\end{equation}
Hence, we do not measure $\widehat{Z}$, but with a sufficiently important number of recorded data points, we can over evaluate it. As a consequence, the photon number distribution $\rho_{nn}$ that we finally obtain will be shifted towards the higher values. The population terms of the low $n$ of $\rho_{nn}$ will be under estimated, and that of the high values of $n$ will be over estimated. Consequently, this distribution will allow us to form an upper bound on the witness. The more the experimental states produced are squeezed on one quadrature, say the $\widehat{X}$ quadrature, the less $\moy{\widehat{Z}}$ will be affected by the value of the other quadrature $\widehat{P}$. As a consequence, with sufficiently squeezed states, $\moy{\widehat{Z}_{sup}}$ will be close to $\moy{\widehat{Z}}$. This will allow us to improve the bound of the witness from Method 1. Note that other ways to form the upper bound of Eq.~\eqref{Zsup} can be envisaged, for instance matching the most important value of $\widehat{P}_2$ with that of $\widehat{X}_1$, and then iterate this procedure. 

\end{document}